\documentclass[useAMS,usenatbib,onecolumn]{mn2e}
\usepackage{graphicx}
\usepackage{amssymb}
\usepackage{lscape}
\usepackage{longtable}
\usepackage{array}
\newcommand\msun{$\rm\,M_\odot$}

\newcommand\mMr{$m_\mathrm{max}$-$M_\mathrm{ecl}$ relation}
\newcommand\mecl{$M_\mathrm{ecl}$}
\newcommand\meclmax{$M_\mathrm{ecl,max}$}
\newcommand\mmax{$m_\mathrm{max}$}
\title[The \mMr, the IMF and IGIMF]{The \mMr, the IMF and IGIMF: probabilistically sampled functions?}
\author[C.~Weidner et al.]
{C.~Weidner$^{1,2}$\thanks{E-mail: cweidner@iac.es}, P.~Kroupa$^{3}$\thanks{E-mail: pavel@astro.uni-bonn.de} and  J.~Pflamm-Altenburg$^{3}$\thanks{E-mail: jpflamm@astro.uni-bonn.de}\\
$^{1}$Instituto de Astrof{\'i}sica de Canarias, Calle V{\'i}a L{\'a}ctea s/n, E38205, La Laguna, Tenerife,
Spain\\
$^{2}$Dept. Astrof{\'i}sica, Universidad de La Laguna (ULL), E-38206 La Laguna, Tenerife, Spain\\
$^{3}$Argelander-Institut f\"ur Astronomie, Auf dem H{\"u}gel 71, D-53121 Bonn, Germany
}

\begin{document}
\bibliographystyle{aa}
\date{Accepted . Received 2013; in original form }

\pagerange{\pageref{firstpage}--\pageref{lastpage}} \pubyear{2013}

\maketitle

\label{firstpage}

\begin{abstract}
We introduce a new method to measure the dispersion of \mmax\,values of star clusters and show that the observed sample of \mmax\,is inconsistent with random sampling from an universal stellar initial mass function (IMF) at a 99.9\% confidence level. The scatter seen in the \mmax-\mecl\,data can be mainly (76\%) understood as being the result of observational uncertainties only. The scatter of \mmax\,values at a given \mecl\,are consistent with mostly measurement uncertainties such that the true (physical) scatter may be very small.

Additionally, new data on the local star-formation regions Taurus-Auriga and L1641 in Orion make stochastically formed stellar populations rather unlikely. The data are however consistent with the local IGIMF (integrated galactic stellar initial mass function) theory according to which a stellar population is a sum of individual star-forming events each of which is described by well defined physical laws. Randomly sampled IMFs and henceforth scale-free star formation seems to be in contradiction to observed reality. 
\end{abstract}



\section{Introduction}
\label{se:intro}
The stellar initial mass function (IMF, $\xi(m)$) describes the distribution of masses of stars, whereby d$N = \xi(m) \mathrm{d}m$ is the number of stars formed in the mass interval $m$, $m + \mathrm{d}m$. It is one of the most important distribution functions in astrophysics as stellar evolution is generally determined by the mass of the stars. The IMF therefore regulates the chemical enrichment history of galaxies, as well as their mass-to-light ratios and influences their dynamical evolution. Theoretically unexpected, the IMF is found to be invariant through a large range of conditions like gas densities and metallicites \citep{Kr01,Kr02,Ch03,EKW08,BCM10,KWP13} and is well described by the canonical IMF (Appendix~\ref{app:IMF}). Though, it has to be kept in mind that often the concept of an universal IMF is understood as a constant slope of the IMF, ignoring the upper and lower mass limits. As the slope (for stellar masses above 1 \msun) has been found to be constant (within the uncertainties) for star clusters in the Milky Way and the Magellanic clouds \citep{Kr02,Mass03}, an invariant IMF is widely used to not only describe individual star clusters but also stellar populations of whole galaxies. But, the question remains whether the IMF, derived from and tested on star cluster scales, is the appropriate stellar distribution function for complex stellar populations like galaxies. In this context, it has emerged that if all the  stars in a galaxy form with a canonical IMF\footnote{With 'form with a canonical IMF' it is meant that the form of the IMF of the star-forming region follows the canonical IMF but the upper mass limit is regulated by the \mMr.} and all these IMFs of all star-forming events (spatially and temporally correlated star formation events/CSFE) are added up the resulting integrated galactic initial mass function of stars (IGIMF) differs substantially from the canonical IMF. It should be pointed out here that the principal concept of the IGIMF - {\it the galaxy-wide IMF (= IGIMF) of a galaxy is always the sum of all star-formation events within a galaxy} - is in any case always true. The ingredients for the IGIMF as applied here are listed as follows:

\begin{itemize}
\item[1.] The IMF, $\xi(m)$, within star clusters is assumed to be canonical (see Appendix~\ref{app:IMF}),
\item[2.] the CSFEs populate an embedded-cluster mass function (ECMF), which is assumed to be a power-law of the form, $\xi_\mathrm{ecl}(M_\mathrm{ecl})$ = d$N$ / d$M_\mathrm{ecl} \propto M_\mathrm{ecl}^{-\beta}$,
\item[3.] the relation between the most-massive star in a cluster, $m_\mathrm{max}$, and the stellar mass of the embedded cluster, $M_\mathrm{ecl}$ \citep{WK04,WK05b,WKB09},
\item[4.] the relation between the star-formation rate (SFR) of a galaxy and the most-massive young ($<$ 10 Myr) star cluster, $\log_{10}(M_\mathrm{ecl, max})$ = $0.746 \times \log_{10}(SFR)$ + 4.93 \citep{WKL04}.
\end{itemize}

Uncertainties are only introduced by the details of star-formation. Properties like the slope of the embedded cluster mass function (ECMF) and its lower mass end ($M_\mathrm{ecl,min}$) in dwarf galaxies or the top/bottom-heaviness of the IMF in the pc-scale star-formation events \citep{MKD10} are examples which need further studies. Therefore any models calculated within the IGIMF-theory contain uncertainties and can not be final. Conversely, it is possible to use observed relations and dependencies of galaxies to refine the understanding of the pc-scale star-formation events within the IGIMF-theory \citep{KWP13}. 

However, the severity of the difference of the IGIMF to the underlying canonical IMF is strongly dependent on how stars form. Two extreme models can be discussed in the context of how stars should be sampled from the IMF. The first model, random sampling, assumes that the IMF is a probability density function. A star cluster is then an ensemble of stars, characterised by its number of stars, $N$, which are randomly drawn from the IMF. Until recently, this approach was used almost exclusively. The second extreme would be optimal sampling \citep{KWP13}. Optimal sampling implies the existence of the \mMr\, \citep{WK05b,WKB09} and populates a star cluster of mass \mecl\,with the optimal number of stars starting from the the most-massive star, \mmax, such that 
no gaps in the distribution arise. This sampling method is based on the notion that the physics of star-formation is self-regulated in a resource-limited environment.

In Section~\ref{se:mmaxmecl} the \mMr\,is presented and discussed, while in Section~\ref{se:2} it is addressed if it is possible to identify the \mMr\,by studying individual star clusters. The expected number of O, B and A stars in star clusters are calculated for an invariant canonical IMF and these are compared with the respective numbers for the canonical IMF with the \mMr. The new data and results for Taurus-Auriga and L1641 are presented in Section~\ref{se:taurus}. Finally, the results are discussed in Section~\ref{se:diss}. Appendix~\ref{app:newstc} lists the young star clusters included in the updated \mMr\,and Appendix~\ref{app:IMF} defines the canonical IMF.

\section{The $\lowercase{m}_\mathrm{\lowercase{max}}$-$M_\mathrm{\lowercase{ecl}}$ relation}
\label{se:mmaxmecl}

\begin{figure}
\begin{center}
\includegraphics[width=8cm]{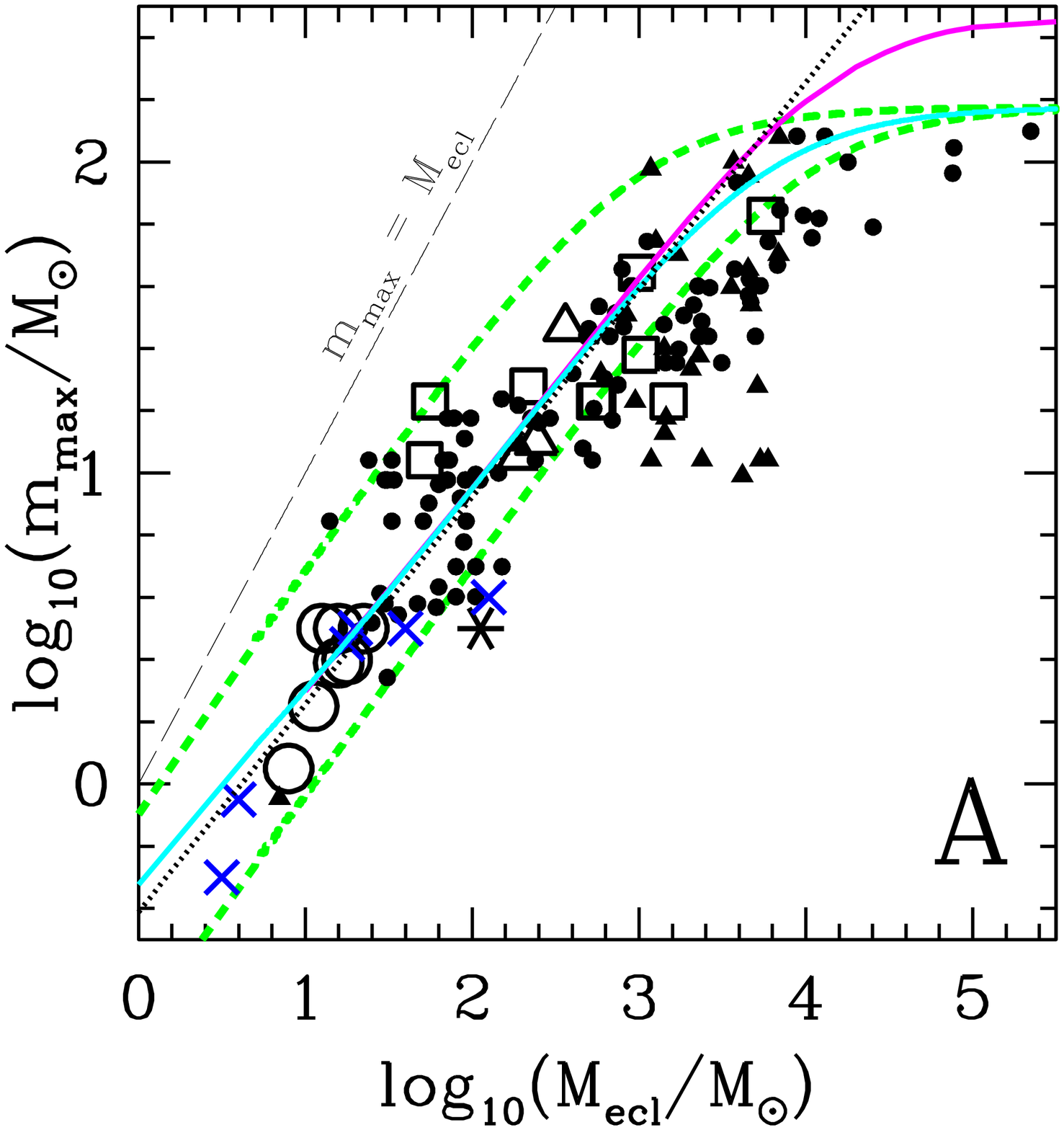}
\includegraphics[width=8cm]{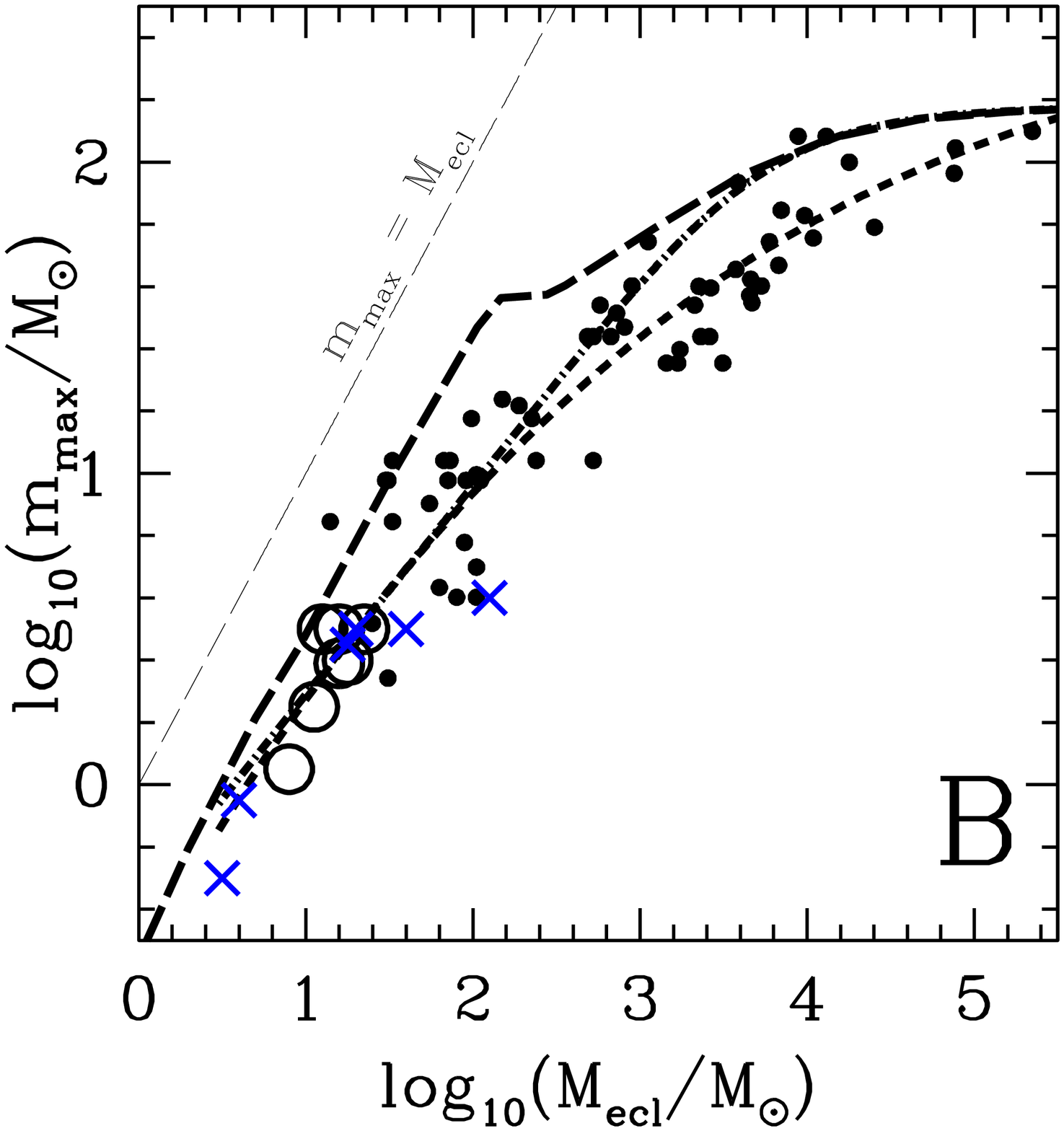}

\vspace*{-2.5cm}

\includegraphics[width=8cm]{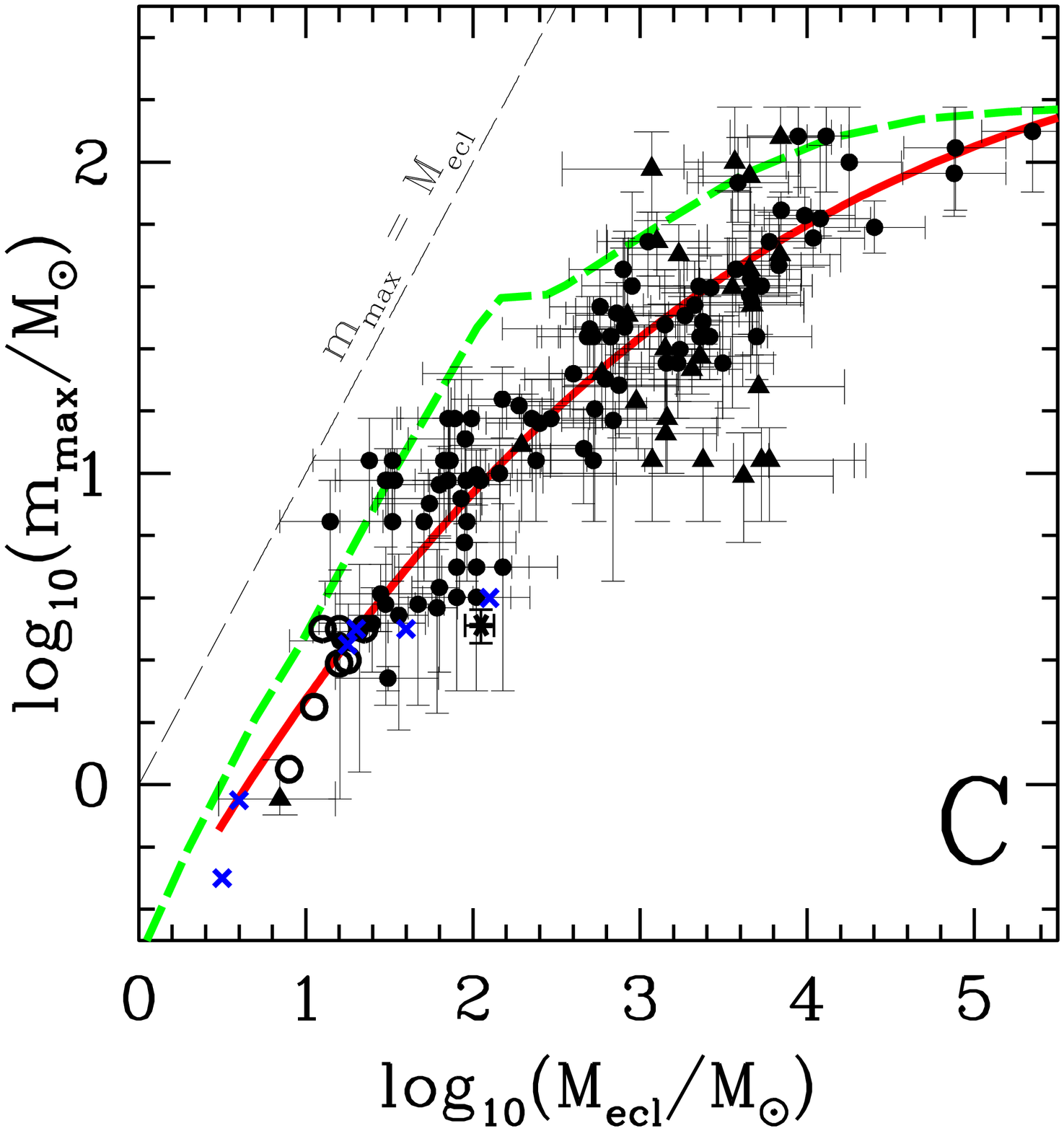}
\vspace*{-1.5cm}
\caption{{\bf Panel A:} The mass of the most-massive star ($m_\mathrm{max}$) in an embedded cluster versus the stellar mass of the young dynamically un-evolved "embedded" cluster ($M_\mathrm{ecl}$). The filled dots are observations compiled by \citet{WKB09}, the filled triangles are new data presented here for the first time (see Appendix~\ref{app:newstc}), the open circles are new data for small clusters in Taurus-Auriga from \citet{KM10}, while the crosses are three other star-forming regions (IC 348, Chameleon I and Lupus 3) discussed in \citet{KM10}. The asterisk symbolises all Taurus-Auriga data of \citet{KM10} combined as if it were a single cluster. Three SPH cluster formation models are indicated by the open triangles \citep{BBV03,SLB09} and the boxes are mm-observations of massive pre-stellar star-forming regions in the Milky Way \citep{JSA09}. The solid lines through the data points are the analytical \mMr\,when using a fundamental upper mass limit, $m_\mathrm{max*}$, of 150 $M_\odot$ (lower light-grey solid line, cyan in the online colour version) and $m_\mathrm{max*}$ = 300 $M_\odot$ (upper dark-grey solid line, magenta in the online colour version). The dashed grey (green in the online colour version) lines are the 1/6 and 5/6th quantiles which would encompass 66\% of the most-massive stars if they were randomly sampled from the IMF. The dotted black line shows the prediction for a relation by \citet{BBV03} from numerical models of relatively low-mass molecular clouds ($\le$ 10000 $M_\odot$). The thin long-dashed line marks the limit where a cluster is made out of one star. {\bf Panel B:} Only the data from panel A are plotted which have uncertainties less than 110\% in \mecl\,and \mmax. These have been used to calculate a 3rd-order polynomial fit (short-dashed line, eq.~\ref{eq:mmaxmecl2}) which is shown together with a fit to the analytical relation (dash-dotted line) as given in \citet{PWK07}. Also shown by the long-dashed line is the mean relation from random sampling, $R$(\mecl), inferred from 10$^7$ Monte-Carlo clusters \citep{WK05b}. Note that the shape of $R$(\mecl) was mathematically confirmed by \citet{SM08}. {\bf Panel C:} Showing the \mmax\,against \mecl\,for the whole sample like in panel A but also showing the errors. 
Like panel A but including the errors in \mmax\,and \mecl. The solid line its the 3rd-order polynomial fit (eq.~\ref{eq:mmaxmecl2}) and the long-dashed line is the mean relation from random sampling, $R$(\mecl).} \label{fig:mmaxmecl}
\end{center}
\end{figure}

The \mMr, as shown in Fig.~\ref{fig:mmaxmecl}, has been analytically presented in \citet{WK04}, observationally established by \citet{WK05b} and refined in \citet{WKB09}, while already briefly theoretically discussed in \citet{R78}. It signifies that the typical upper mass limit to which the IMF is sampled, $m_\mathrm{max}$, changes systematically with the stellar mass of the cluster, $M_\mathrm{ecl}$, the stars have formed in and is incompatible with a scale-free IMF. Note that in Fig.~\ref{fig:mmaxmecl} several new objects are added which have not been published before (see Appendix~\ref{app:newstc} for the full list of clusters and their properties).

For the clusters for which the number of stars above a mass limit or within a mass range are given in the literature, the cluster mass, \mecl, is calculated by assuming a canonical IMF (Appendix~\ref{app:IMF}) from 0.01 to 150 \msun\,and extrapolating to the total population from the observational mass limits. The observed number of stars and its mass limits are given in Tab.~\ref{tab:newclusters}. For the error determination of \mecl, the error in the number of stars (if given in the literature) is combined with the assumption that all stars could be unresolved binaries for an upper mass limit and that 50\% of the stars are misidentified as cluster members for the lower mass limit. In the cases where no observed numbers of stars and their mass limits were given in the references, literature values for \mecl\,have been used.

The mass of the most-massive star, \mmax, is either deduced from the spectral type of the most-massive star by using a spectral-type--stellar-mass relation for O stars \citep{WV10} and B stars \citep{HHC97} or, when this was not possible and in the case of exotic spectral types (like Luminous Blue Variables or Wolf-Rayet stars), literature values have been inserted into the table. For the errors in \mmax, $\pm$ 0.5 was assumed for the spectral subclass. Panel C of Fig.~\ref{fig:mmaxmecl} shows the errors in \mmax\,and \mecl.

We emphasise here that we have been using {\it all} available data on very young populations and that the selection criteria are {\it only} one of age being younger than 4 Myr and no supernova remnants being present in the cluster. That is, we have not been discarding any data. 

Note that the smoothed particle hydrodynamical (SPH) simulations \citep*[i.e.][]{BBV03} and grid based computations of star formation with the FLASH code \citep{PBK11} show good agreement with the empirical \mMr\,\citep{WKB09,KWP13}. This suggests that the formation of stars within the cloud cores is mostly driven by growth processes in a medium with limited resources. \citet{PBK11} refer to this process as {\it fragmentation induced star formation}.

\subsection{Challenges for the $\lowercase{m}_\mathrm{\lowercase{max}}$-$M_\mathrm{\lowercase{ecl}}$ relation}
\label{sub:challenges}

The existence of a \mMr\,is not without challenge. \citet{PG07} and \citet{MC08} discuss a list of relatively young A and B stars around which \citet{TPP97,TPN98,TPN99} had searched for star clusters. The majority of these clusters are not included in the \citet{WKB09} study as a) they  are either too old ($>$~4 Myr for 25 of 35 objects) or they are b) gas-free. A strict age limit is necessary because of the short life time of massive stars. For older clusters it is impossible to determine if they have had more massive stars which had exploded as supernovae. Furthermore, such clusters loose considerable amounts of stars due to stellar dynamical processes, even before gas expulsion \citep{LMD84,Go97,KAH01,GB06,PMH06b,BG06,PAK06,WKNS06,WLB07,WBM10}. For the same reason completely gas-free objects are unsuited as gas-expulsion removes large numbers of stars and therefore reduces the mass of the cluster, \mecl, significantly \citep{KAH01,WKNS06,WBM10}. Furthermore, the Testi-sample does not include any study of the proper motions or radial velocities and the A and B stars and clusters are only searched for in a 0.2 to 1.0 pc radius around the stars. Any even only slowly ejected/evaporated A or B stars would therefore be seen as being isolated. A detailed study of the kinematics of the Testi-sample is currently underway. The Testi-sample has in common four objects with the near infra-red study of young star-forming regions by \citet{WL07} and which are included in the \citet{WKB09} study. Additionally, the \citet{MC08} study changes and adapts their own sample until they arrived at an acceptable probability for their hypothesis that the data is consistent with random sampling. Taking their full sample, \citet{MC08} perform a KS-test and arrive at a probability for random sampling of $10^{-17}$! Only after removing the Testi-sample and some other clusters they arrive at a 20\% probability for random sampling. This reduction of the sample is justified by \citet{MC08} by arguing that clusters around G type stars are missing as "{\it young stars of close to solar mass and below have not been systematically targeted for surrounding clusters}" but as is evident from fig.~2 of \citet{MC08}, what are actually missing are clusters around Ae stars ($\approx$ 2.5\msun, their fig.~2 does not even extend below 1\msun) with several hundred stars. Exactly what Testi had been looking for. Also they only use the low-mass end (\mecl\,$\lesssim$ 1200\msun) of the \citet{WK05b} sample and merely conclude that ''the data are not indicating any striking deviation from the expectations of random drawing''. This does not rule out an other mechanism than random sampling but merely states that the low-mass clusters they studied do not allow for any discrimination whether or not there is a \mMr, especially as no other hypothesis was tested. \citet{WKB09} showed for clusters with \mecl\,between 100 and 1000\msun, while the percentage of \mmax\,in the 1/6th and 5/6th quantiles is compatible with random sampling, that the data has only a probability of 1.9 $\times$ 10$^{-7}$ for being symmetrically distributed around the median, thus making random sampling highly unlikely. For more massive clusters (\mecl\,$>$ 1000\msun), the probability for symmetry around the median is even lower (2.8 $\times$ 10$^{-9}$).

Furthermore, one has to keep in mind that star clusters do not have identical initial conditions. For very similar masses, initial differences in metallicity, rotation, magnetic field strength and orientation and star-formation efficiency of the giant molecular cloud will lead to some spread of the mass of the most-massive star even if there were to exist an exact \mMr\,under ideal conditions such as no rotation of the cluster forming cloud, identical boundary conditions and chemical composition \citep{KWP13}. Also binary stellar evolution can alter the mass of the most-massive object \citep{BB05,DLI11}. Finally, but perhaps most importantly, the data suffer from significant observational errors (50\% in $m_\mathrm{max}$ and $M_{\rm ecl}$) such that much of the dispersion seen in Fig.~\ref{fig:mmaxmecl} may be due to measurement uncertainties. Even with all these potential sources for variation of the \mMr\,it is surprising that 77\% of the clusters with errors in Table~\ref{tab:newclusters} {\it are} compatible within their errors with being form one universal \mMr.

Further criticism of the \mMr~comes from the claim by some workers that massive stars can form in isolation. The study by \citet{DTP04} and \citet{DTP05} arrived at an upper limit fraction of 4 $\pm$ 2 \% of known O stars as candidates for the formation of massive stars in isolation. Unfortunately, this number is usually but falsely used as the percentage of O stars that definitely formed in isolation. One example would be: \citet{KCK10} write "{\it \citet{DTP04,DTP05} find that 4\% $\pm$ 2\% of galactic O stars formed outside of a cluster of significant mass, which is consistent with the models presented here [...], but not with the proposed cluster-stellar mass correlation}". \citet{LOW10}, \citet{BVG11} and \citet{SHG11} propose a handful of apparently formed-in-isolation O stars in the Magellanic Clouds with similar arguments. The thorough study by \citet{GWKP12} has eliminated any statistically significant evidence for the existence of O stars formed in isolation. The remaining candidates are likely two-step-ejections \citep[a massive binary is ejected and the more massive component explodes as a supernova, changing direction and velocity of the secondary;][]{PK10}. Two-step ejections must be common as the vast majority of field and runaway O stars are in binaries \citep{CHN12}. Further studies are currently underway to discuss the new samples of 'candidates' for the formation of massive stars in isolation presented by \citet{BBE12} and \citet{OLK13}.

Recently, it has been claimed that modelling of observations of young star clusters in the starburst dwarf galaxy NGC 4214 disproves the \mMr\,\citep{ACC13}. This claim is disproven in Weidner et al. (2013, submitted). The main issue of \citet{ACC13} is to assume that the \mMr\,is a fixed truncation limit instead of the mean of the observations.\\

Also, it is important to keep in mind that all current observations of young star-forming regions, including the surface-density profiles, are in agreement with stars forming in embedded clusters when properly taking into account stellar dynamical processes which result in dissolution of star clusters and allowing for more than one cluster to be formed in a given molecular cloud \citep{PKO12}.

Note that random sampling is only given by using a number $N$ of stars, taking these randomly from the IMF and calculating the \mmax\,and \mecl\,for each $N$. Choosing a \mecl\,and filling it randomly with stars is mass-constrained sampling and results in a completely different expected \mMr\,as mass-constrained sampling changes the IMF \citep{WK05b}. This is because it is not possible to reach in this way \mecl\,at a 100\% level but only with a certain precision and because, especially for low-mass clusters, it happens that a star randomly drawn is either more massive than the cluster itself or adding it to the cluster changes the mass significantly. Generally, such stars are discarded and therefore the IMF changes. Mass-constrained sampling can therefore never be scale-free.

\subsection{Statistical tests}
\label{sub:stat}
For easy implementation of the observed \mMr\,and in order to calculate statistical tests a 3rd-order polynomial fit is calculated,

\begin{equation}
\label{eq:mmaxmecl2}
y = \left\{ \begin{array}{ll}
a_0 + a_1 x + a_2  x^2 + a_3  x^3&\textrm{for}\,3 \le M_\mathrm{ecl}/M_\odot \le 2.5 \cdot 10^5\\
\log_{10}(150/M_\odot)&\textrm{for}\,M_\mathrm{ecl} > 2.5 \cdot 10^5 M_\odot
\end{array} \right .
\end{equation}
with $y$ = $\log_{10}(m_\mathrm{max}/M_\odot)$, $x$ = $\log_{10}(M_\mathrm{ecl}/M_\odot)$, $a_0$ = -0.66 $\pm$ 0.18, $a_1$ = 1.08 $\pm$ 0.22, $a_2$ = -0.150 $\pm$ 0.075, and $a_3$ = 0.0084 $\pm$ 0.0078. The correlation factor $R^2$ is 0.91. The fit is only valid for 3 $\le$ $M_\mathrm{ecl}/M_\odot$ $\le$ $2.5 \cdot 10^5$. In panel B of Fig.~\ref{fig:mmaxmecl} only clusters with uncertainties lower than 110\% in \mecl\footnote{The \mecl\,is usually calculated by extrapolating from an observed number of stars to the total sample by assuming an IMF (see also Appendix~\ref{app:newstc}). For the upper and lower end of \mecl\,it is  assumed that all observed stars could be unresolved binaries or 50\% of them could be foreground/background contamination. This usually results in errors of about 100\% for \mecl} and \mmax\,have been used to obtain the fits. 77\% of the clusters which have errors in \mecl\,and \mmax\,in Table~\ref{tab:newclusters} are fully consistent within their errors with this fit. It needs to be kept in mind that the uncertainties listed in Table~\ref{tab:newclusters} only address unresolved binaries and potentially misidentified stars. Other error sources, like variable extinction, stellar variability, star loss due to gas expulsion and dynamical interactions \citep{OK12} are not taken into account. It is therefore not unlikely that all the scatter seen in Fig.~\ref{fig:mmaxmecl} is due to observational uncertainties and not to variations of \mmax\,between similar clusters. This means that the impact of physical parameters like rotational velocities of the stars, binary stellar evolution, metallicity and magnetic fields may be very small.

The important question remains whether or not this expanded sample of most-massive stars in star-forming regions is compatible with random sampling of stars from the IMF or not. The answer to this questions has important implications for the theory of star-formation. If random sampling of stars from the IMF is observed then the isolated formation of O stars would be possible. If O stars can only form in denser environments, massive star formation would be a distinct process like, e.g.~competitive accretion \citep{BBV03} or fragmentation induced star formation \citep{PBK11}, different from that of low-mass stars.

In order to quantify if the observed \mMr\,is in agreement with random sampling or not, we perform the statistical tests. To do so, the geometrical distances of the observed sample of \mmax-\mecl\,tuples to mean \mMr\,from random sampling\footnote{Note that random sampling is only given by using a number $N_\mathrm{ecl}$ of stars in a cluster, taking these randomly from the IMF and calculating the \mmax\,and \mecl\,values for each $N_\mathrm{ecl}$. Starting from a given \mecl\,and populating it with stars randomly drawn from the IMF is mass-constrained sampling and results in a completely different expected \mMr\,as mass-constrained sampling changes the IMF \citep{WK05b}. Mass-constrained sampling can therefore never be scale-free.}, $R(M_\mathrm{ecl})$. The $R(M_\mathrm{ecl})$ was calculated from a large sample of Monte-Carlo experiments in \citet{WK05b}, which itself agrees well with a mathematical determination by \citet{SM08}. 

\begin{figure}
\begin{center}
\includegraphics[width=8cm]{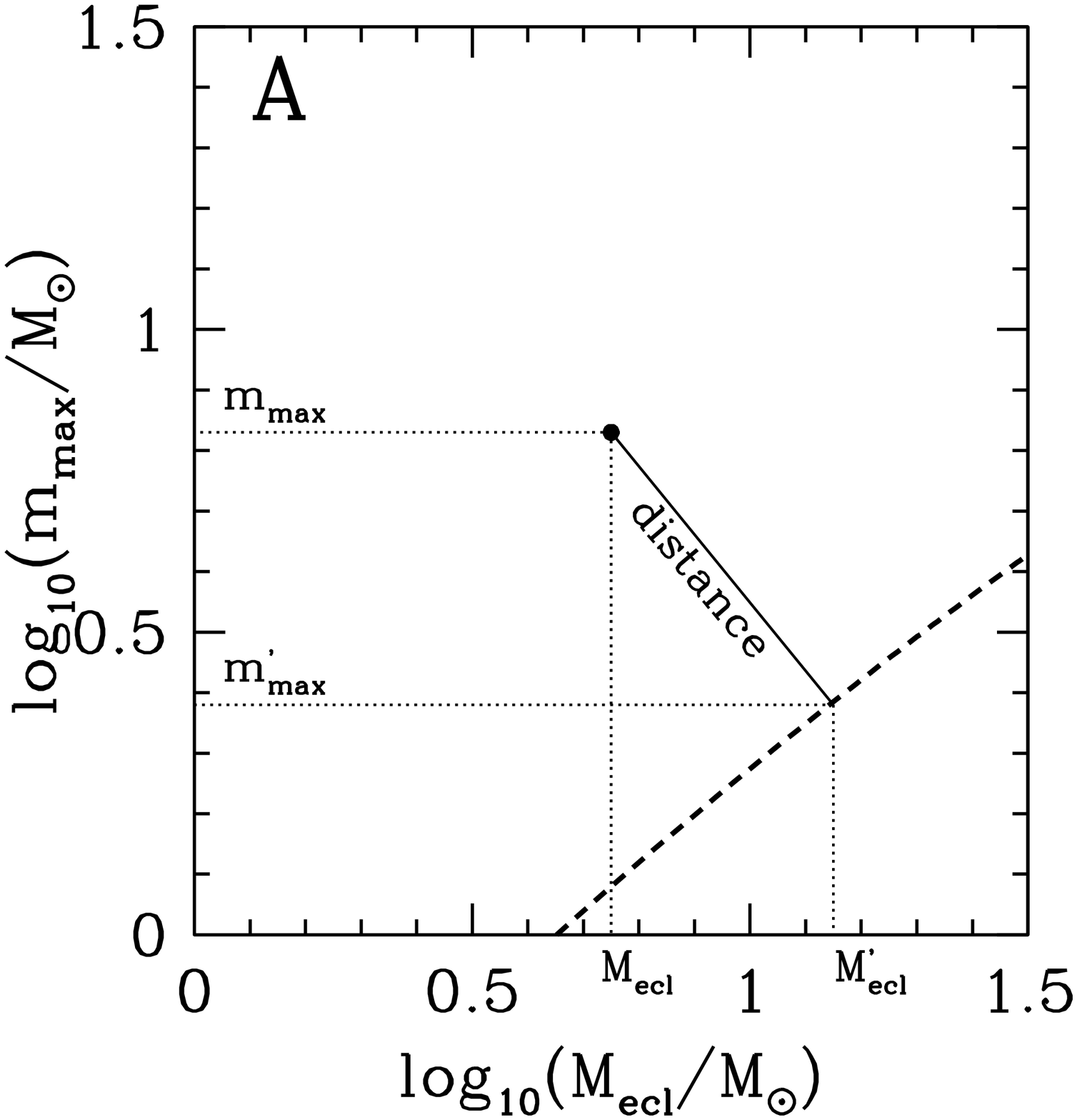}
\includegraphics[width=8cm]{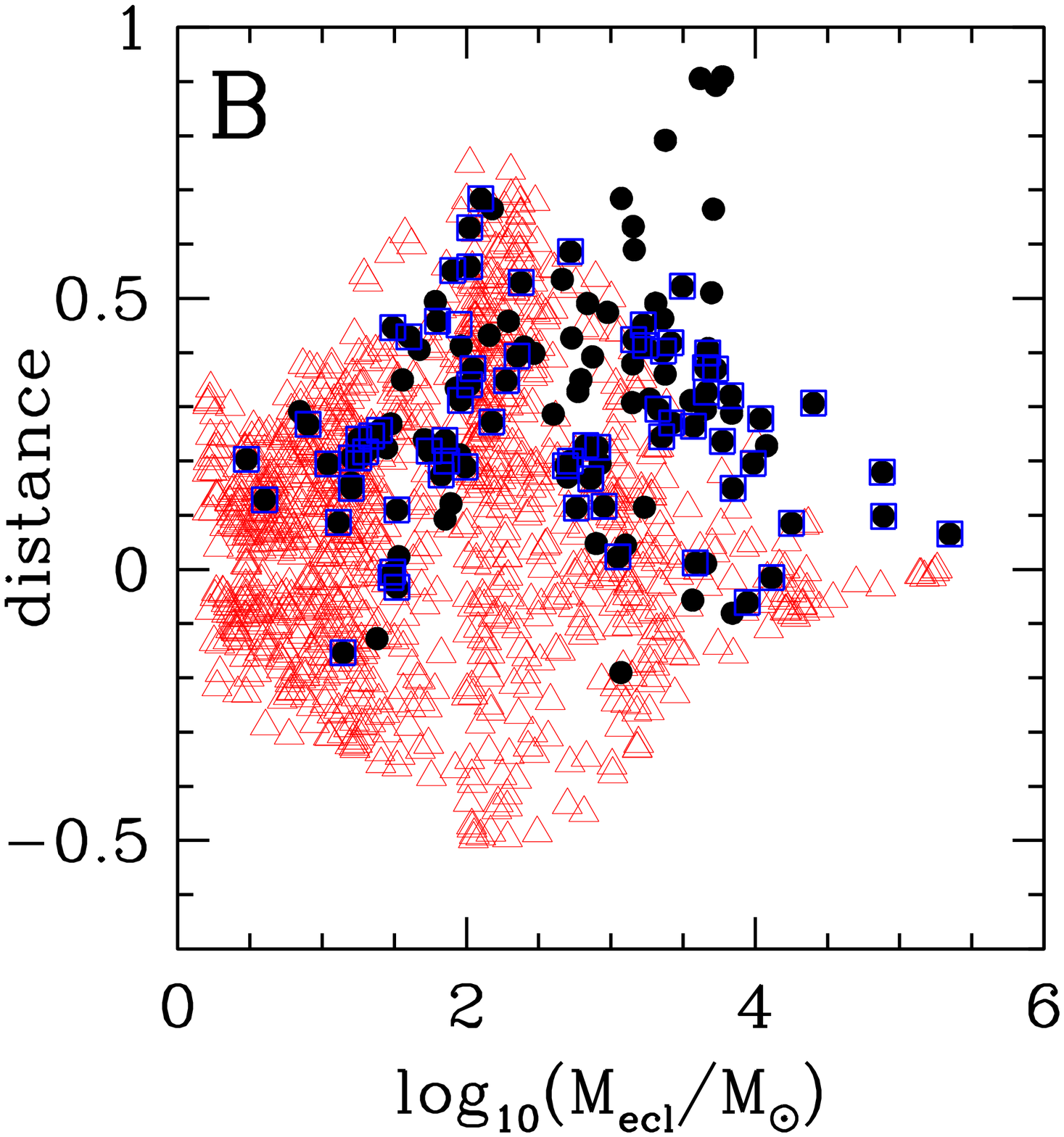}
\vspace*{-1.5cm}
\caption{{\bf Panel A:} Visualisation of how the distances between the curves and the data are calculated. The dot symbolises one data point, the dashed line is the expectation for random sampling, $R(M_\mathrm{ecl})$, the solid line is the distance between dot and dashed line and the dotted lines indicate the coordinates used to calculate the distance. {\bf Panel B:} Geometric distances of the observational data and the Monte-Carlo clusters from the expectation for random sampling. The filled dots compare the full observational sample to the mean expected from random sampling, the open (blue) boxes the low-error observational sample (panel B of Fig.~\ref{fig:mmaxmecl}) and the (red) triangles the Monte Carlo clusters. Note that only a sub-sample of the 10$^6$ Monte-Carlo clusters is plotted for clarity.} \label{fig:dist}
\end{center}
\end{figure}

The geometrical distances between the data and the expectation of random sampling, $R$(\mecl), are calculated by determining the shortest distance for each data point $i$ as follows,

\begin{equation}
\label{eq:dist}
{\rm distance}_i = \min \left(\sqrt{[\log_{10}(m_\mathrm{max, i}) - \log_{10}(m^{'}_\mathrm{max})]^2 + [\log_{10}(M_\mathrm{ecl, i}) - \log_{10}(M^{'}_\mathrm{ecl})]^2}\right),
\end{equation}
where $m_\mathrm{max, i}$ and $M_\mathrm{ecl, i}$ is the \mmax\,and the $M_\mathrm{ecl}$ of the i-th data point and $m^{'}_\mathrm{max}$ and $M^{'}_\mathrm{ecl}$ are, respectively, the \mmax\,and the $M_\mathrm{ecl}$ values of the curves. When the observed \mmax\,is lower than the $R$(\mecl) for the given \mecl\,the distance is multiplied with minus one. Panel A of Fig.~\ref{fig:dist} shows a symbolic representation of this process. 

These distances are then compared via a KS-test to the distances of a large sample of numerically created clusters to $R(M_\mathrm{ecl})$. These numerically derived clusters are build the following way. A number of stars, $N_\mathrm{ecl}$, is chosen and randomly filled with stars from the IMF. The stars are added up the obtain the \mecl\,and the most-massive one, \mmax, is searched for. This procedure is repeated in total for 10$^6$ clusters. The $N_\mathrm{ecl}$ are randomly drawn from an embedded cluster number function\footnote{An embedded cluster number function is almost identical to the embedded cluster mass function (ECMF) but instead of providing the mass for clusters chosen from the function it provides the number of stars, $N_\mathrm{ecl}$, for a cluster chosen from it. The slope of both functions are the same but they are offset to each other by the mean number of stars for a given IMF.}. The embedded cluster number function, $\xi_\mathrm{ecl}(N_\mathrm{ecl})$, gives the number of stars $N_\mathrm{ecl}$ in the interval $N_\mathrm{ecl}$, $N_\mathrm{ecl}+\mathrm{d}N_\mathrm{ecl}$, is $\xi_\mathrm{ecl}(N_\mathrm{ecl}) \propto N^{-\beta}_\mathrm{ecl}$. Here as parameters are a slope $\beta$ of 2 from $N_\mathrm{ecl, min}$ = 10 stars to $N_\mathrm{ecl, max}$ = 10$^6$ stars.

Note that taking the \mecl\,directly from the ECMF makes random sampling challenging. Depending on how the \mecl\,is filled with stars, it is generally impossible to exactly reach \mecl\,and when stars are rejected (e.g.~when the star itself is much more massive than the star cluster) the IMF is changed. Therefore, this process should be referred to as mass-constrained sampling and not random sampling. The differences between several sampling methods are discussed in \citet{WK05b}.

The distances of the observed sample and the Monte-Carlo sample to $R(M_\mathrm{ecl})$ are plotted in panel B of Fig.~\ref{fig:dist}. The filled dots in the plot mark the distances of the full observational sample to $R$(\mecl), the open (blue) boxes are the distances of the low-error observational sample of panel B in Fig.~\ref{fig:mmaxmecl} to $R$(\mecl) and the (red) triangles are the distances of a selection of 1116 of the 10$^6$ Monte-Carlo generated clusters to $R$(\mecl). It is clearly visible that the distances of the Monte-Carlo clusters are well distributed around the expected \mMr\,for random sampling while the observations are not.

\begin{figure}
\begin{center}
\includegraphics[width=8cm]{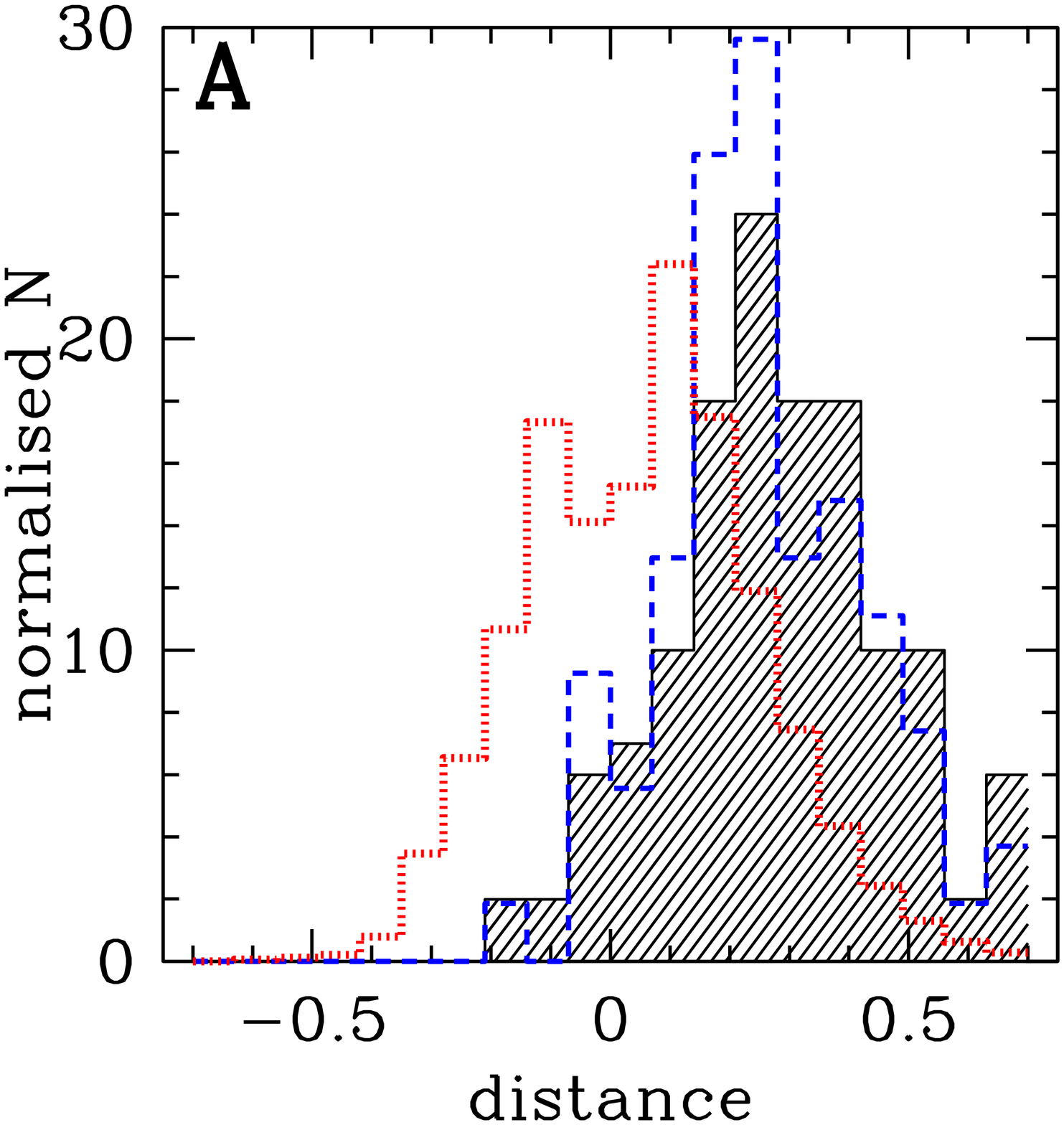}
\includegraphics[width=8cm]{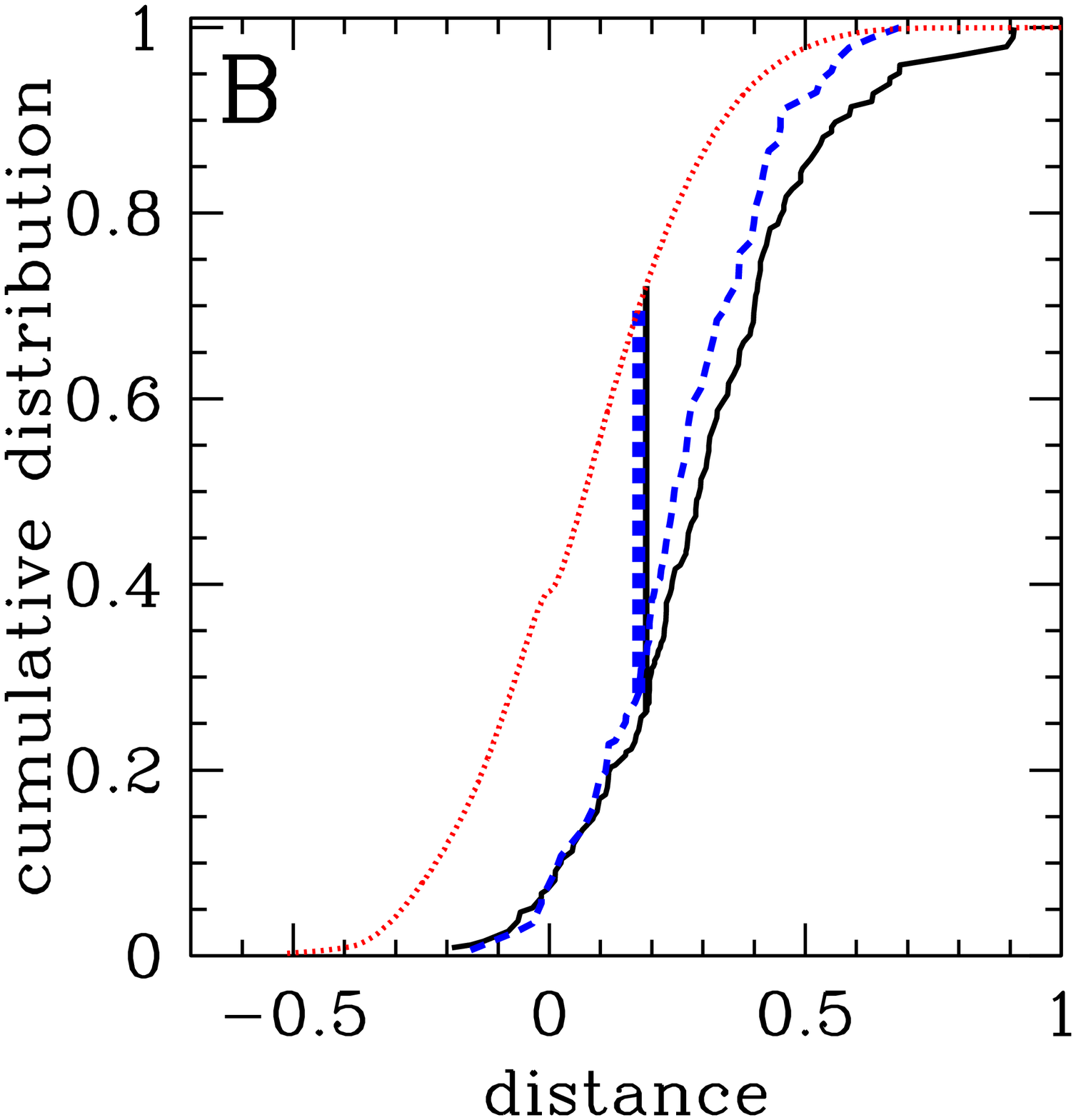}
\vspace*{-1.5cm}
\caption{{\bf Panel A:} Histograms of geometric distances of panel B of Fig.~\ref{fig:dist}, normalised to the total number of observed clusters of 137. The solid (black) shaded histogram shows the distances of the full observational sample to the mean of random sampling, while the (blue) dashed-line plots the distances to the mean of random sampling for the reduced (low-error) observational sample. The dotted (red) is the histogram of 10$^6$ Monte-Carlo clusters to the mean of random sampling. {\bf Panel B:} The cumulative distributions of the distances for the KS-test. The solid (black) line shows the cumulative distribution of the distances of the full observational sample to the mean of random sampling and the same is plotted for the reduced observational sample by the dashed (blue) line. The dotted (red) line is cumulative distribution of the distances of 10$^6$ Monte-Carlo clusters to the mean of random sampling. The vertical lines indicate the points of largest separation, $D$, between cumulative distribution of the Monte-Carlo clusters and the full observational sample (solid vertical line) and the reduced one (dashed vertical line).} \label{fig:expdist}
\end{center}
\end{figure}

KS-tests are preformed on the data shown in panel B of Fig.~\ref{fig:dist} for the two hypothesis A) that the distances between the full sample of observations to $R$(\mecl) are from the same distribution as the distances for the 10$^6$ Monte-Carlo clusters to $R$(\mecl) and B) that the distances of the reduced observational sample of panel B in Fig.~\ref{fig:mmaxmecl} to $R$(\mecl) are from the same distribution as the distances for the 10$^6$ Monte-Carlo clusters to $R$(\mecl). The $D$-value\footnote{When comparing two samples by means of a KS-test the critical value $D_{\alpha}$ for $\alpha$ = 0.001 (for sample sizes larger than 12) is then arrived at by $\sqrt{(n+n^{'})/(n*n^{'})}$ times 1.95 = 0.167 for $n$ = 137 (observational data) and $n^{'}$ = 10$^6$ (Monte-Carlo data). If $D$ is larger than this $D_{\alpha}$, the given hypothesis can be rejected at a 99.9\% confidence level \citep{LW01}.} for hypothesis A is 0.483. The critical $D_{\alpha}$ value for two samples with $n$ = 137 data points and $n^{'}$ = 10$^6$ data points and $\alpha$ = 0.001 is 0.167.  As $D > D_\alpha$, hypothesis A is rejected at the 99.9\% confidence level. The two samples are not from the same distribution. For hypothesis B the $D$-value is 0.440 - significantly larger than $D_{\alpha = 0.001}$ = 0.227 (for $n$ = 74 and $n^{'}$ = 10$^6$). Therefore, hypothesis B can also be rejected at a 99.9\% confidence level. This leads to the conclusion with very high significance that the \mmax\,values of the observed sample of clusters are not randomly drawn from the IMF and that the observed \mMr\,is not compatible with random sampling. The cumulative distributions for the KS-tests are shown in panel B of Fig.~\ref{fig:expdist}.

To test a possible impact of the choice of slope of the cluster mass function, the same Monte-Carlo experiment as before has been repeated by drawing again 10$^6$ clusters but using cluster number function slopes of $\beta$ = 1.6 and $\beta$ = 2.4. After calculating the expectation values for random sampling for these two slopes the distances of the full observational sample and the Monte-Carlo samples to $R$(\mecl, $\beta$) have been calculated and KS-tests performed. The $D$ for the distances of the $\beta$ = 1.6 Monte-Carlo experiment to $R(M_\mathrm{ecl}, \beta = 1.6)$ and the distances of the complete observed sample to $R(M_\mathrm{ecl}, \beta = 1.6)$ is $D$ = 0.417, which is significantly larger than $D_{\alpha = 0.001}$ = 0.167. For the $\beta$ = 2.4 Monte-Carlo sample $D$ = 0.546 $>$ $D_{\alpha = 0.001}$ = 0.167. It is therefore save to conclude that observations rule out scale-free star-formation.

\section{The expected number of massive stars from random sampling and from the $\lowercase{m}_\mathrm{\lowercase{max}}$-$M_\mathrm{\lowercase{ecl}}$ relation}
\label{se:2}

It is important to constrain the assumptions that enter the IGIMF theory. An important assumption is the existence of the \mMr. A possible test of this relation is to assess whether or not the expected number of high-mass stars in a star cluster differs significantly when the clusters are sampled from an invariant canonical IMF with a fixed upper mass limit or when using the \mMr, or even when optimal sampling \citep{KWP13} is assumed.

Using the canonical IMF (Appendix~\ref{app:IMF}), $\xi(m)$, it is possible to calculate the total number of stars in a cluster,
\begin{equation}
\label{eq:Ntot}
N_\mathrm{tot} = \int_{m_\mathrm{min}}^{m_\mathrm{max}} \xi(m) dm = k \int_{m_\mathrm{min}}^{m_\mathrm{max}} \xi^{'}(m) dm,
\end{equation}
when given a lower mass limit, $m_\mathrm{min}$, an upper mass limit, $m_\mathrm{max}$ and the normalisation constant, $k$ (eq.~\ref{eq:4pow}). The $m_\mathrm{min}$ is set to 0.1 $M_\odot$ and $m_\mathrm{max}$ is either given by the fundamental upper mass limit for stars, which is thought to be $m_\mathrm{max *}= $150 $M_\odot$ \citep[][but see \citealt{BK12}]{WK04,OC05,Fi05,Ko06} or the empirical \mMr\,(eq.~\ref{eq:mmaxmecl2}). The normalisation constant, $k$, is not known and can be derived by setting $M_\mathrm{ecl}$ which is the physically relevant quantity,

\begin{equation}
 \label{eq:Mtot}
M_\mathrm{ecl} = k \int_{m_\mathrm{min}}^{m_\mathrm{max}} m\,\xi^{'}(m) dm.
\end{equation}

\noindent
The expected number of stars in different mass bins $[m_{1},m_{2}]$ is then
\begin{equation}
\label{eq:N}
N_{\ast} = \int_{m_1}^{m_2} \xi(m) dm.
\end{equation}
For the  calculations in this paper, A stars are assumed to have masses between 1.65 to 3 $M_\odot$ \citep{A04}, while O stars are defined with masses from 18 to 150 $M_\odot$ \citep{WV10}. Stars with masses between 3 and 18 $M_\odot$ are set as B stars.

For five different cluster masses the expected numbers of A, B and O stars are listed in Table~\ref{tab:numbers150} for a fixed upper mass limit of 150 $M_\odot$, three different ages (1, 10 and 100 Myr) and random sampling from the canonical IMF. For the same cluster masses but using eq.~\ref{eq:mmaxmecl2} to determine \mmax\,the numbers are in Table~\ref{tab:numbersvar}. In both tables only the numbers of stars are shown ($m >$ 0.1 $M_\odot$). Brown dwarfs are not included. 

The 1 $\sigma$ errors included in the tables are determined by a Monte-Carlo method. For each mass, 10000 clusters are generated with masses deviating at most $\pm$ 10\% from the target masses and either randomly sampling $m$ from the IMF with an upper mass limit of 150 $M_\odot$ or by using sorted sampling \citep{WK05b} in order to introduce the  \mMr. For sorted sampling $M_\mathrm{ecl}$ is divided by the mean mass of the used IMF (for the canonical IMF between 0.1 and 150 $M_\odot$ it is $m_\mathrm{mean} =$ 0.61 $M_\odot$) which results in the expected number of stars for the cluster. This number of stars is randomly taken from the IMF, sorted by mass and added, starting from the lowest mass star. When $\sum_{i=1}^{N}m_{i} = M_\mathrm{ecl} \pm 10\%$, $m_\mathrm{max} = m_\mathrm{N}$, or, when $\sum_{i=1}^{N-1}m_{i} = M_\mathrm{ecl} \pm 10\%$ then $m_\mathrm{max} = m_\mathrm{N-1}$. If the cluster mass is not reached, an additional number of stars is randomly chosen and sorted into the first list of stars. The additional number is determined by dividing the difference of the target cluster mass and the sum of the first star list by the mean mass. This extended list of stars is then summed up again in the same manner as before. This procedure is iterated until the cluster mass is reached to a tolerance of 10\%. For each of the resulting distributions of O, B and A stars the half-full-width-half-maximum is then calculated and used as 1 $\sigma$ errors.

As can be seen in the Tables~\ref{tab:numbers150} and \ref{tab:numbersvar} the choice of the different upper mass limit has very little effect on the expected numbers of A, B and O stars even within the uncertainties. Introducing a cluster mass dependent upper mass increases the total number of stars expected in low-mass clusters but the change is too small to be measurable. Testing for the existence of a  \mMr\,is thus not possible using the number of massive stars for a given $M_\mathrm{ecl}$. Instead, the distribution of $m_\mathrm{max}$ values for different $M_\mathrm{ecl}$ needs to be considered as has been done in \S~\ref{se:mmaxmecl}.

\begin{table*}
\caption{\label{tab:numbers150} The expected numbers of A, B and O stars for individual embedded star-formation events of stellar mass $M_\mathrm{ecl}$ with different ages, an invariant canonical IMF sampled randomly and  $m_\mathrm{max*}$ = 150 $M_\odot$. To derive the 1 $\sigma$ errors, $N_\mathrm{tot}$ is used to randomly realise 10000 clusters with masses within $\pm$ 10\% of $M_\mathrm{ecl}$. There are no O stars at 100 Myr.}
\begin{tabular}{ccccccccccc}
$M_\mathrm{ecl}$ &$N_\mathrm{tot}$&$m_\mathrm{max*}$&A$_{1 \mathrm{Myr}}$&B$_{1 \mathrm{Myr}}$&O$_{1 \mathrm{Myr}}$&A$_{10 \mathrm{Myr}}$&B$_{10 \mathrm{Myr}}$&O$_{10 \mathrm{Myr}}$&A$_{100 \mathrm{Myr}}$&B$_{100 \mathrm{Myr}}$\\
$M_\odot$&$\ge$ 0.1 $M_\odot$&$M_\odot$&&&&\\
\hline
10& 16&150&1 $\pm$ 1& 0 -0/+1&0 $\pm$ 0& 1 $\pm$ 1& 0 -0/+1&0 $\pm$ 0&1 $\pm$ 1& 0 -0/+1\\
100& 164&150&5 $\pm$ 3& 4 $\pm$ 2& 0 -0/+1& 5 $\pm$ 3& 4 $\pm$ 2& 0 -0/+1& 5 $\pm$ 3& 2 $\pm$ 2\\
1000& 1636&150&48 $\pm$ 8& 35 $\pm$ 7& 3 $\pm$ 2& 48 $\pm$ 8&35 $\pm$ 7&1 $\pm$ 1& 48 $\pm$ 8&21 $\pm$ 6\\
10000& 16359&150&477 $\pm$ 23& 350 $\pm$ 20& 32 $\pm$ 6&477 $\pm$ 23&350 $\pm$ 20&7 $\pm$ 3& 477 $\pm$ 23&190 $\pm$ 20\\
100000 & 163587&150&4767 $\pm$ 83& 3498 $\pm$ 72& 322 $\pm$ 20&4767 $\pm$ 83&3498 $\pm$ 72&64 $\pm$ 10& 4767 $\pm$ 83&2139 $\pm$ 49\\
\hline
\end{tabular}
\end{table*}

\begin{table*}
\caption{\label{tab:numbersvar} Like Table~\ref{tab:numbers150} but assuming the empirical eq.~\ref{eq:mmaxmecl2} instead of a fixed upper mass limit of 150 $M_\odot$ and using sorted sampling \citep{WK05b} which is very close to optimal sampling \citep{KWP13}.}
\begin{tabular}{ccccccccccc}
$M_\mathrm{ecl}$ &$N_\mathrm{tot}$&$m_\mathrm{max}$&A$_{1 \mathrm{Myr}}$&B$_{1 \mathrm{Myr}}$&O$_{1 \mathrm{Myr}}$&A$_{10 \mathrm{Myr}}$&B$_{10 \mathrm{Myr}}$&O$_{10 \mathrm{Myr}}$&A$_{100 \mathrm{Myr}}$&B$_{100 \mathrm{Myr}}$\\
$M_\odot$&$\ge$ 0.1 $M_\odot$&$M_\odot$&&&&&\\
\hline
10& 23& 2.3&1 -1/+0& 1 -1/+0&0 $\pm$ 0& 1 -1/+0& 1 -1/+0&0 $\pm$ 0&1 -1/+0& 1 -1/+0\\
100& 185& 10.8&5 $\pm$ 2& 4 $\pm$ 1& 1 -1/+0& 5 $\pm$ 2& 4 $\pm$ 1& 1 -1/+0& 5 $\pm$ 2& 2 -2/+1\\
1000& 1704& 44.3&48 $\pm$ 7& 37 $\pm$ 7& 3 $\pm$ 1& 48 $\pm$ 7&37 $\pm$ 7&1 -1/+0& 48 $\pm$ 7&19 $\pm$ 4\\
10000& 16562& 101.4&476 $\pm$ 25& 354 $\pm$ 18& 32 $\pm$ 5&476 $\pm$ 25&354 $\pm$ 18&7 $\pm$ 3& 476 $\pm$ 25&194 $\pm$ 17\\
100000 & 164210& 135.2&4794 $\pm$ 95&3514 $\pm$ 56&324 $\pm$ 20&4794 $\pm$ 95&3514 $\pm$ 56&65 $\pm$ 9& 4794 $\pm$ 95&2148 $\pm$ 50\\
\hline
\end{tabular}
\end{table*}

\subsection{The number of stars in Taurus-Auriga and in the Orion cloud L1641}
\label{se:taurus}

The Taurus-Auriga region is of special interest because it is the nearest known site of star-formation. Currently, 352 young stellar objects (YSO) and stars are known to be situated in about 8 small clusters each with a radius of $\approx$ 0.3 pc \citep{KM10} covering about 50 times 50 pc. 42 of the YSOs have masses above 1 $M_\odot$ but the most-massive object weighs only 3.25 $M_\odot$ \citep{KM10}. If star-formation is following random sampling, it should not matter if any substructure exists within the Taurus-Auriga region and the most-massive star should be set by the total number of stars \citep{Elme00b}. For a canonical IMF and random sampling 9 stars above 3.25 $M_\odot$ are expected to be present for a cluster with 42 YSOs above 1 $M_\odot$. To estimate the likelihood that a cluster with so many stars above 1 $M_\odot$ has no stars above 3.25 $M_\odot$ a Monte-Carlo experiment is used. 10$^6$ clusters are randomly filled with stars until each has 42 stars above 1 $M_\odot$. Of these 10$^6$ clusters only 58 have no stars above 3.25 $M_\odot$. Thus the probability is $\approx$ $6 \cdot 10^{-5}$ for this to occur. If all the 42 YSOs are unresolved binaries, then for a flat mass-ratio distribution from random sampling \citep{K08c}, 31 YSOs of the 42 systems are expected to be more massive than 1 $M_\odot$ and 6 should be above 3.25 $M_\odot$. The probability of observing no such stars is $\approx$ $1.3 \cdot 10^{-3}$.

A possible interpretation of this result is that the IMF in Taurus-Auriga is different from the canonical IMF. On the other hand \citet{KM10} found that the eight most-massive stars of the eight sub-structures in Taurus-Auriga follow very well the  \mMr\,(Fig.~\ref{fig:mmaxmecl}). But when combining the mass of all the small clusters into one cluster moves Taurus-Auriga outside the 66\% region such that it lies well below the $m_\mathrm{max}$-$M_\mathrm{ecl}$-relation as is shown by the asterisk in Fig.~\ref{fig:mmaxmecl}. \citet{KM10} showed that these sub-clusters are compatible with the canonical IMF, but the whole region is not a single 'cluster' but a conglomeration of several small clusters \citep{KB03}. The deficiency of stars above 1 $M_\odot$ for the whole region is exactly a sign of a "local IGIMF effect": a number of low-mass clusters with a low upper stellar mass limit which together constitute a stellar population but without massive stars. This has also been found in simulations of collapsing molecular clouds \citep{MCB10}. The notion that Taurus-Auriga is not one but many small clusters would then also explain the so-called 'inverse mass segregation' found for the whole field \citep{PBG10,MC11}. Furthermore, it is important to bear in mind when studying the Taurus-Auriga region that here several young populations are found with ages from below 1 Myr up to 30 or 40 Myr \citep{GBA07}. The original molecular clouds in which the stars older than $\approx$ 10 Myr formed have long since dispersed and these stars are now well mixed with the younger populations. \citet{NTS97} found that the radial velocity dispersion of the stars in Taurus-Auriga agrees very well with the stars of about a few $10^7$ yr to have traveled about 50 pc and are therefore well mixed with the younger populations in the region.

In a recent study \citet{HHA12} found a similar result for Orion. While the L1641 cloud has a similar total number of YSOs as the Orion Nebula Cluster (ONC), the cloud has no O or B stars - in disagreement to what would be expected form random sampling  from an invariant IMF. \citet{HHA12} find 2362 stars above $\approx$ 0.1 $M_\odot$ in L1641 and the most-massive star to have 16 $M_\odot$. The probability for no stars above 16 $M_\odot$ to have formed from a sample of 2362 stars is $4 \times 10^{-3}$.  Therefore, {\it random or stochastic sampling as a description of star formation in molecular clouds is ruled out with a very high confidence of 99.6\%}. In \citet{HHA13} the authors come to the similar conclusion that it is highly unlikely that the stars in the ONC and L1641 are drawn from the same population. Interestingly, after dividing L1641 in a northern (L1641-n) and a southern part (L1641-s) \citet{HHA13} estimate the total stellar mass L1641-s with $\approx$ 1000 \msun, the extend of the star-formation period to 10 Myr and a \mmax\, of 7 \msun. This would result in an average star-formation rate (SFR) of 10$^{-4}$ $M_\odot$ yr$^{-1}$. The relation between the SFR and the most-massive cluster, \meclmax, in a galaxy from \citet{WKL04} predicts \meclmax\,= 88 $M_\odot$. The \mMr\,results in a \mmax\,for such a cluster of about 8 \msun, which is within the observational uncertainties of \mmax\,in L1641-s. If optimal sampling \citep{KWP13}\footnote{The software McLuster has been used to calculate the optimally sampled distributions \citep{KMK11}.} is used to distribute the total mass of L1641-s with a cluster mass function (with a slope of $\beta$ = 2) between 5 \msun\,and 88 \msun, 64 clusters are to be expected. Within the area of L1641-s, it is probably not possible to separate all these clusters. The same calculation can be done for Taurus-Auriga. \citet{KM10} list 8 sub-clusters with a total mass of 112 \msun. Again when using a star-formation period of 10 Myr, the SFR-\meclmax-relation gives a \meclmax\,of 17.2 \msun, which if fairly close to the observed \meclmax\,in Taurus-Auriga of 22 \msun. The \mMr\,gives 2.8 \msun\,as \mmax, again very close to the observed 3.25 \msun\,(see Table~\ref{tab:newclusters}). Optimal sampling results in 10 clusters between 8 and 17.2 \msun\,for Taurus-Auriga while 8 are observed. If one would use 7.5 Myr instead of 10 Myr, the calculated numbers would represent the observed ones even better. L1641-s and Taurus-Auriga can therefore readily be seen as good examples of the IGIMF on local scales and that molecular clouds tend to produce clusters of star clusters. Table~\ref{tab:ligimf} sums up the predictions and observations for Taurus-Auriga and L1641-s.

\begin{table*}
\caption{\label{tab:ligimf} Observed and predicted quantities for Taurus-Auriga and L1641-s.}
\begin{tabular}{ccccccc}
Name&predicted&observed&predicted&observed&predicted&observed\\
& \meclmax& \meclmax&$N_\mathrm{clusters}$&$N_\mathrm{clusters}$& \mmax& \mmax\\
&[\msun]&[\msun]&&&[\msun]&[\msun]\\
\hline
Taurus-Auriga&17.2&22&10&8&2.8&3.2\\
L1641-s&88&-&64&-&8&7\\
\hline
\end{tabular}
\end{table*}

Note that a similar situation appears to have been found in the G305 star-forming complex \citep{FTH12} which includes the two young star clusters Danks 1 and Danks 2 \citep{DCT12}. But the available literature data does not allow for a quantitative analyses yet.

\section{Conclusions}
\label{se:diss}

Whether or not star-formation is a purely stochastic process is of profound importance for the understanding of stellar populations and for the chemical evolution of galaxies. The existence of a non-trivial \mMr\,would be in contradiction to random star-formation. Several criticisms of the \mMr\,have been published in the past \citep{PG07,MC08,LOW10,BVG11,SHG11}, but as can be seen in \S~\ref{sub:challenges} these can be proven as being based on misconceptions or misinterpretations of the available data. Apparent 'proofs' for the formation of O stars in isolation were shown to be questionable in \citet{GWKP12}. An updated list of star clusters used to determine the \mMr\,is presented in Table~\ref{tab:newclusters} in Appendix~\ref{app:newstc}. This data is also used to determine a 3rd-order polynomial fit (eq.~\ref{eq:mmaxmecl2}) which is plotted in panel B of Fig.~\ref{fig:mmaxmecl}. The expanded sample of clusters is tested against the expectation for random sampling of stars from the IMF, and the hypothesis that stars are randomly drawn from the IMF can be excluded at a 99.9\% confidence level. Furthermore, 76\% of clusters with error measurements (127 clusters of Table~\ref{tab:newclusters}) lie within their errors on the 3rd-order polynomial fit to the data. Therefore, there is little room for an intrinsic dispersion of the \mMr\,with physical parameters like metallicity, magnetic fields, rotation or binary stellar evolution.

Further, we calculated the expected numbers of O, B and A stars in star clusters of different mass (Tables~\ref{tab:numbers150} and \ref{tab:numbersvar}) when assuming random sampling or when the most-massive star is constrained by the \mMr. It turns out that the numbers do not depend on whether a physical \mMr\,(Table~\ref{tab:numbersvar}) is assumed to be valid or not (Table~\ref{tab:numbers150}). But the existence of such a relation can be addressed by studying whole star-forming regions. One example is Taurus-Auriga, where random sampling from the canonical IMF can not explain the relatively low mass of the most-massive star while the local IGIMF theory accounts excellently for the observations. The sub-clusters individually follow the \mMr\,but the most-massive sub-cluster sets the most-massive star for the whole region. When assuming that the total mass of all sub-clusters formed over a time scale of 10 Myr and using the SFR-\meclmax\,relation as well as the \mMr, the most-massive star in Taurus-Auriga as well as the most-massive sub-cluster and the number of sub-clusters agree well between theory and observations (Table~\ref{tab:ligimf}). Another such case is the Orion molecular cloud, where essentially all OB stars are situated within the ONC, but the whole cloud has produced 5000-10000 stars in total. As the ONC is already deficient in OB stars \citep{PAK06}, it becomes quite clear that again random sampling would not account for the stellar census. In \citet{HHA12} and \citet{HHA13} it is found that the part of the cloud south of the ONC, which is called L1641, is deficient in O stars on a three to four $\sigma$ level. Again the local IGIMF theory predicts the mass of the most-massive star in the region well. Thus, star-forming molecular clouds may be the best testbed for discerning whether star formation initially samples stars randomly from the IMF or whether the non-invariant IGIMF theory is the relevant description. Comparing the number of sub-clusters and their most-massive stars of resolvable star forming regions with the local IGIMF theory would allow to put further constrains on the local IGIMF theory and the \mMr. Though, one has to keep in mind that observing sub-clusters can be very difficult as these tend to disperse quickly and projection effects of the real 3D space structure can prevent an identification in many cases. As Taurus-Auriga is very nearby and a relatively low-density star-forming region the study of the sub-clusters is easier.

While still widely used in cosmology and extragalactic stellar populations studies, an invariant IMF for galaxies is  challenged by mounting observational evidence \citep{LGB05,VD07,HG07,Da07,E08,WHT08,MWK09,LGT09,HAJ10,GHS10,DKP12,CMA12,FBR13}. The results presented here and in application to extra-galactic problems \citep{KWK05,PWK07,PAK08,PWK09,RCK09,WK07a} clearly demonstrate that the IGIMF-theory is readily able to reproduce the observational data very well. The IGIMF-theory, based on the knowledge of the local star-formation process, is therefore a useful description of large-scale star-formation in whole galaxies.

Independently of the observational results but adopting the observed correlations and distribution functions in star-forming galaxies, it has become evident that the IMF of a whole galaxy, the IGIMF, must differ from the canonical IMF, thus implying the IGIMF to vary with the SFR of the galaxy. While the above demonstrates that the IGIMF theory is in good agreement with the latest observational data it is necessary to further test the theory. Here we have argued that nearby star-forming regions are fully compatible with the assumptions that enter the IGIMF theory and that they falsify random sampling.

Our conclusions are that randomly sampled IMFs are most-likely in contradiction to the observed reality, that is, a purely stochastic descriptions of star formation on the scales of a pc and above are deemed to be highly unlikely. Instead, star formation seems to follow well defined laws. Even if one assumes that the here studied clusters are a small sub-sample of the possibilities of star-formation it needs to be kept in mind that for scale-free star-formation, also any sub-sample has to be immediately scale-free as well. Therefore using cluster and OB associations to test the nature of star-formation is perfectly valid. If star formation were to be inherently stochastic, in the sense that stars are randomly selected from the full IMF, then the star-formation simulation results (see \S~\ref{se:mmaxmecl}) would imply that a strong randomisation agent during star formation is necessary. This is because the simulations already lead to a good agreement with the empirical \mMr\,despite being based mostly on gravitational physics. In other words, the well-ordered process of stars arising from a molecular cloud core captured by pure gravitationally driven accretion would have to be upset completely through this putative agent. A possibility for such an agent might be stellar feedback but it is difficult to see how feedback could unsettle the whole molecular cloud quickly enough to make star formation appear to be a random process. The concept that star-formation is random can therefore be considered to being probably unphysical.


\section*{Acknowledgements}
This study has made use of the SIMBAD database, operated at CDS, Strasbourg. This work has been supported by the Programa Nacional de Astronom{\'i}a y Astrof{\'i}sica of the Spanish Ministry of Science and Innovation under grant AYA2010-21322-C03-02.

\begin{appendix}
\section{Star cluster catalogue}
\label{app:newstc}

Table~\ref{tab:newclusters} shows the clusters used for the \mMr\,analysis in this work and plotted in Fig.~\ref{fig:mmaxmecl}. Besides the objects already listed in \citet{WKB09}, 26 new clusters are added and one was corrected (No.~85 in this list). Furthermore, the sub-clusters as characterised by \citet{KM10} for Taurus Auriga, IC 348 and Cha I are included as individual objects. The references for the data are shown in Table~\ref{tab:newclustersref}.

For the clusters for which the number of stars above a mass limit or within a mass range are given in the literature, the cluster mass, \mecl, is calculated by assuming a canonical IMF (Appendix~\ref{app:IMF}) from 0.01 to 150 \msun\,and extrapolating to the total population from the observational mass limits. For the error determination, the error in the number of stars (if given in the literature) is combined with the assumption that all stars could be unresolved binaries for an upper mass limit and that 50\% of the stars are misidentified as cluster members for the lower mass limit. In the cases where no observed numbers of stars and their mass limits were given in the references, literature values for \mecl\,have been used.

The mass of the most-massive star, \mmax, is either deduced from the spectral type of the most-massive star by using a spectral-type--stellar-mass relation for O stars \citep{WV10} and B stars \citep{HHC97} or, when this was not possible and in the case of exotic spectral types (like Luminous Blue Variables or Wolf-Rayet stars), literature values have been inserted into the table. For the errors in \mmax, $\pm$ 0.5 was assumed for the spectral subclass. For example, an O5V star would be evaluated as O4.5V and O5.5V for the maximum and minimum mass, respectively. In general, the \mmax\,values changed a few percent (maximal 10\% in very few cases) for most of the O stars compared to the ones published in \citet{WKB09}, as in that work a preliminary version of the spectral-type--stellar-mass relation of \citet{WV10} had been used.

For a cluster to be included in the Table, it has to fulfil several criteria. Most importantly, the age has to be below 4 and better below 3 Myr to exclude the possibility that the most-massive star has already exploded as a supernova. Clusters with known supernova remnants have been excluded even if their age would formally be below 4 Myr. Binary stellar evolution and subsequent stellar merging can lead to premature supernovae. Preferably, the cluster should be still at least partly enshrouded in its molecular cloud. Gas expulsion leads to loss of stars and while it is unlikely to loose the most-massive star through this process \citep{OK12} it will bias \mecl\,towards lower masses.

\begin{landscape}
\begin{longtable}{ccccccccc}
\caption{\label{tab:newclusters} Literature data for the \mMr. The table shows empirical cluster masses ($M_\mathrm{ecl}$), maximal star masses ($m_\mathrm{max, obs}$) within these clusters, cluster ages (age), distances (D), the numbers of stars above or within certain mass limits (in $M_\odot$), the name and the spectral type of the most-massive star. References for the data are given in Table~\ref{tab:newclustersref}.}\\
\hline
No.&Designation&$M_{\rm ecl}$&$m_\mathrm{max~obs}$&age & D& \# of stars&Id $m_{\rm  max}$&Sp Type\\
&&[$M_{\odot}$]&[$M_{\odot}$]&[Myr]&[kpc]&$>$ \msun&&\mmax\\
\hline
\endfirsthead
\caption{continued.}\\
\hline
No.&Designation&$M_{\rm ecl}$&$m_\mathrm{max~obs}$&age & D& \# of stars&Id $m_{\rm  max}$&Sp Type\\
&&[$M_{\odot}$]&[$M_{\odot}$]&[Myr]&[kpc]&$>$ \msun&&\mmax\\
\hline
\endhead
\hline
\endfoot
1&IC 348 2$^{\ast,\dagger}$& 3&0.5&1.3&0.31&-&-&-\\
2&Cha I 3$^{\ast,\dagger}$& 4 &0.9&2.0&0.17&-&-&-\\
3&B59$^{\ast}$&$7_{-4}^{+8}$&0.9$_{-0.1}^{+0.3}$&2.6 $\pm$ 0.8&0.13&20 $\ge$ 0.1&-&-\\
4&Taurus-Auriga 7$^{\ast,\dagger}$&8&1.1&1-2&0.14&-&-&-\\
5&Taurus-Auriga 8$^{\ast,\dagger}$&8&1.1&1-2&0.14&-&-&-\\
6&Taurus-Auriga 6$^{\ast,\dagger}$&11&1.8&1-2&0.14&-&-&-\\
7&Taurus-Auriga 3$^{\ast,\dagger}$&13&3.0&1-2&0.14&-&-&-\\
8&IRAS 05274+3345$^{\dagger}$&14$_{-7}^{+15}$&7.0 $\pm$ 2.5&1.0&1.8&15 $>$ 0.24&-&B2\\
9&Taurus-Auriga 5$^{\ast,\dagger}$&16&2.5&1-2&0.14&-&-&-\\
10&Mol 139&16 $\pm$ 8&2.9 $\pm$ 2.0&$<$1&7.3&-&-&-\\
11&Taurus-Auriga 2$^{\ast,\dagger}$&16&3.0&1-2&0.14&-&-&-\\
12&Taurus-Auriga 4$^{\ast,\dagger}$&18&2.5&1-2&0.14&-&-&-\\
13&Lupus 3$^{\ast,\dagger}$&18&2.8&2.0&0.2&-&-&-\\
14&Cha I 2$^{\ast,\dagger}$& 20 &3.0&2.0&0.17&-&-&-\\
15&Mol 143&21 $\pm$ 10&3.1 $\pm$ 2.0&$<$1&5.0&-&-&-\\
16&Taurus-Auriga 1$^{\ast,\dagger}$&22&3.0&1-2&0.14&-&-&-\\
17&IRAS 06308+0402&24$_{-13}^{+25}$&11.0 $\pm$ 4.0&1.0&1.6&16 $>$ 0.37&-&B0.5\\
18&VV Ser$^{\dagger}$&25$_{-13}^{+27}$&3.3 $\pm$ 1.0&0.6&0.44&24 $>$ 0.3&VV Ser&B9e\\
19&VY Mon&28$_{-15}^{+29}$&4.1 $\pm$ 1.0&0.1&0.8&26 $>$ 0.3&VY Mon&B8e\\
20&Mol 8A&30 $\pm$ 15&3.8 $\pm$ 2.0&$<$1&11.5&-&-&-\\
21&IRAS 05377+3548$^{\dagger}$&30$_{-15}^{+32}$&9.5 $\pm$ 2.5&1.0&1.8&31 $>$ 0.24&-&B1\\
22&Ser SVS2$^{\dagger}$&31$_{-16}^{+31}$&2.2 $\pm$ 0.2&2.0&0.259 $\pm$ 0.037&50 $>$ 0.17&BD 01$^\circ$ 3689&A0\\
23&IRAS 05553+1631$^{\dagger}$&31$_{ -16}^{+33}$&9.5 $\pm$ 2.5&1.0&2.0&28 $>$ 0.28&-&B1\\
24&IRAS 05490+2658$^{\dagger}$&33$_{-17}^{+36}$&7.0 $\pm$ 2.5&1.0&2.1&30 $>$ 0.29&-&B2\\
25&IRAS 03064+5638$^{\dagger}$&33$_{-17}^{+36}$&11.0 $\pm$ 4.0&1.0&2.2&27 $>$ 0.31&-&B0.5\\
26&IRAS 06155+2319&34$_{ -18}^{+35}$&9.5 $\pm$ 2.5&1.0&1.6&38 $>$ 0.21&-&B1\\
27&Mol 50&36 $\pm$ 18&3.5 $\pm$ 2.0&$<$1&4.9&-&-&-\\
28&Cha I 1$^{\ast,\dagger}$& 40 &3.0&2.0&0.17&-&-&-\\
29&Mol 11&47 $\pm$ 20&3.8 $\pm$ 2.0&$<$1&2.1&-&-&-\\
30&IRAS 06058+2138&51$_{-27}^{+54}$&7.0 $\pm$ 2.5&1.0&2.0&26 $>$ 0.49&-&B2\\
31&NGC 2023$^{\dagger}$&55$_{-28}^{+58}$&8.0 $\pm$ 2.0&3.0&0.4&21 $>$ 0.6&HD 37903&B1.5V\\ 
32&Mol 3&61 $\pm$ 20&3.7 $\pm$ 2.0&$<$1&2.17&-&-&-\\
33&Mol 160$^{\dagger}$&63 $\pm$ 20&4.3 $\pm$ 2.0&$<$1&5.0&-&-&-\\
34&NGC 7129&63$_{-33}^{+104}$&9.2 $\pm$ 3.0&0.1&1.0&53 $>$ 0.3 / 3 $>$ 3&BD 65$^\circ$ 1637&B3e\\
35&IRAS 06068+2030$^{\dagger}$&67$_{-35}^{+70}$&11.0 $\pm$ 4.0&1.0&2.0&59 $>$ 0.28&-&B0.5\\
36&IRAS 00494+5617$^{\dagger}$&71$_{-37}^{+74}$&9.5 $\pm$ 2.5&1.0&2.2&58 $>$ 0.31&-&B1\\
37&V921 Sco&71$_{-36}^{+429}$&15.0 $\pm$ 5.0&0.1-1&0.8&33 $>$ 0.5&V921 Sco&B0e\\
38&IRAS 05197+3355&72$_{ -38}^{+75}$&11.0 $\pm$ 4.0&1.0&3.2&34 $>$ 0.50&-&B0.5\\
39&IRAS 05375+3540$^{\dagger}$&73$_{ -38}^{+78}$&11.0 $\pm$ 4.0&1.0&1.8&74 $>$ 0.24&-&B0.5\\
40&IRAS 02593+6016&78$_{ -41}^{+81}$&15.0 $\pm$ 5.0&1.0&2.2&61 $>$ 0.31&-&B0\\
41&Mol 103$^{\dagger}$&80 $\pm$ 20&4.0 $\pm$ 2.0&$<$1&4.1&-&-&-\\
42&NGC 2071&80$_{-44}^{+89}$&4.0 $\pm$ 2.0&1.0&0.4&105 $>$ 0.2&V1380 Ori&B5\\
43&Cha I (whole field)&80$_{-46}^{+91}$&5.0 $\pm$ 3.0&2.0&0.17 $\pm$ 0.01&237 $>$ 0.04&HD 96675&B6IV/V\\
44&MWC 297&85 $\pm$60&8.3$_{-1.3}^{+13.7}$&0.1-1&0.25 or 0.45&24 $>$ 0.3&-&B1.5V/O9e\\
45&BD 40$^\circ$ 4124&90$_{-49}^{+106}$&12.9$_{-6.0}^{+ 2.0}$&0.1-6&1.0&74 $>$ 0.3/ 3 $>$ 3&BD 40$^\circ$ 4124&B2e\\
46&$\rho$ Oph$^{\dagger}$&91$_{-46}^{+93}$&9.5 $\pm$ 2.5&0.1-1&0.13 $\pm$ 0.02&78 $>$0.3&$\rho$ Oph&BIV\\
47&IRAS 06056+2131&92$_{-49}^{+97}$&7.0 $\pm$ 2.5&1.0&2.0&85 $>$ 0.28&-&B2\\
48&IRAS 05100+3723$^{\dagger}$&98$_{-51}^{+103}$&15.0 $\pm$ 5.0&1.0&2.6&63 $>$ 0.38&-&B0\\
49&R CrA$^{\dagger}$&105$_{-55}^{+114}$&4.0 $\pm$ 2.0&1.0&0.13&55 $>$ 0.5&R CrA&A5eII\\ 
50&NGC 1333$^{\dagger}$&105$_{-54}^{+111}$&5.0 $\pm$ 1.0&1-3&0.25&134 $>$ 0.2&SSV 13&-\\ 
51&Mol 28$^{\dagger}$&105 $\pm$ 20&9.9 $\pm$ 2.0&$<$1&4.5&-&-&-\\
52&IRAS 02575+6017$^{\dagger}$&111$_{-57}^{+116}$&9.5 $\pm$ 2.5&1.0&2.2&91 $>$ 0.31&-&B1\\
53&Taurus-Auriga (whole field)&112 $\pm$ 22&3.2 $\pm$ 0.4&1-2&0.14&-&-  &B1\\
54&IC 348 1$^{\ast,\dagger}$& 126&4.0 $\pm$ 2.0&1.3&0.31&-&BD 31$^{\circ}$643&B5V\\
55&W40&144$_{-80}^{+576}$& 10.0 $\pm$ 5.0&1-2&0.6&3 $>$ 4&IRS 2a&-\\
56&$\sigma$ Ori$^{\dagger}$&150$_{-76}^{+155}$&18.6$_{-5.6}^{+5.4}$&2.5&0.36 $\pm$ 0.06&140 $\pm$ 10 (0.2-1.0)&$\sigma$ Ori A&O9-9.5V\\
57&NGC 2068&151$_{-86}^{+169}$&5.0 $\pm$ 3.0&1.0&0.4&192 $>$ 0.2&HD 38563A&B4V\\ 
58&NGC 2384$^{\dagger}$&189$_{-95}^{+192}$&16.5 $\pm$ 1.5&1.0&2.1&7 $>$ 3&HD 58509&B0.5III\\
59&LkH$\alpha$ 101$^{\ast}$&195$_{-123}^{+295}$&12.3$_{-5.3}^{+5.7}$&1&0.51 $\pm$ 0.1&271 $\ge$ 0.1 $M_\odot$&-&B0-1V\\
60&Mon R2$^{\dagger}$&225$_{-117}^{+236}$&15.0 $\pm$ 5.0&0-3&0.83 $\pm$ 0.05&309 $>$ 0.15&IRS 1SW&B0\\
61&IRAS 06073+1249$^{\dagger}$&239$_{ -120}^{+242}$&11.0 $\pm$ 4.0&1.0&4.8&25 $>$ 1.47&-&B0.5\\
62&Trumpler 24&251$_{ -131}^{+291}$&14.5 $\pm$ 2.5&1.0&1.14&4 $>$ 5&GSC 7872-1609&WN\\
63&IC 5146&293$_{ -226}^{+305}$&15.0 $\pm$ 5.0&1.0&0.9&238 $>$ 0.3 / 5 $>$ 3&BD 46$^\circ$ 3474&B0e\\
64&HD 52266&400 $\pm$ 350&20.9$_{ -6.9}^{+8.1}$&$<$3.0&1.7 $\pm$ 1.0&4 $\pm$ 2 $>$ 4 &HD 52266&O8-9V\\
65&HD 57682&400 $\pm$ 350&20.9$_{ -6.9}^{+8.1}$&$<$3.0&?&4 $\pm$ 5 $>$ 4&HD 57682&O8-9V\\
66&Alicante 5&461$_{ -234}^{+516}$&12.0 $\pm$ 4.0&$<$3.0&3.6$_{-0.4}^{+0.6}$&22 $>$
2.5&A47&B0.7V\\
67&Cep OB3b$^{\dagger}$&485$_{ -243}^{+497}$&27.5$_{ -7.5}^{+9.5}$&3.0&0.8 $\pm$ 0.1&12 $>$ 4&HD 217086&O7Vn\\
68&HD 153426&500 $\pm$ 350&29.1$_{ -7.1}^{+7.9}$&$<$3.0&?&5 $\pm$ 4 $>$ 4&HD 153426&O6.5-7V\\
69&Sh2-294$^{\dagger}$&525$_{ -267}^{+540}$&11.0 $\pm$ 4.0&4.0&3.2&155 $>$ 0.7&S294B0.5V&B0.5V\\
70&NGC 2264$^{\dagger}$&525$_{ -267}^{+537}$&27.5$_{ -7.5}^{+9.5}$&3.0&0.76 $\pm$ 100&1000 $>$ 0.08&S Mon / HD 47839&O7Ve\\
71&RCW 116B&536$_{ -276}^{+557}$&16.1$_{ -9.1}^{+12.9}$&2.5&1.1&102 $>$ 0.95&-&O8V-B1V\\
72&Alicante 1$^{\dagger}$&577$_{ -290}^{+583}$&34.7$_{ -10.7}^{+13.3}$&2-3&4.0 $\pm$ 400&38 $>$ 2.0&BD 56$^\circ$ 864&O6V\\
73&RCW 36$^{\ast}$&591$_{ -305}^{+619}$&20.9$_{ -6.9}^{+8.1}$& 2.5 $\pm$ 0.5& 0.7& 349 $\ge$ 0.04 $M_\odot$&-& O8-9V\\
74&HD 52533&621$_{ -417}^{+1077}$&20.1$_{ -6.1}^{+5.9}$&$<$3.0&?&15 $\pm$ 5 $>$ 4&HD 52533&O8.5-9V\\
75&Sh2-128&666$_{ -342}^{+736}$&27.5$_{ -7.5}^{+9.5}$&2.0&9.4&7 $>$ 7&-& O7V\\
76&NGC 6383$^{\dagger}$&668$_{ -334}^{+671}$&27.5$_{ -7.5}^{+9.5}$&2.0&1.3 $\pm$ 100&21 $>$ 3&HD 159176&O7V + O7V\\
77&NGC 2024&690$_{ -350}^{+706}$&14.8$_{ -10.5}^{+14.2}$&0.5&0.4&309 $>$ 0.5&IRS 2b&O8V - B2V\\
78&HD 195592$^{\dagger}$&725$_{ -364}^{+757}$&32.7$_{ -8.7}^{+8.3}$&$<$3.0&?&18 $\pm$ 3 $>$ 4&HD 195592&O6-6.5V\\
79&Sh2-173&748$_{ -395}^{+901}$&19.2$_{ -6.2}^{+6.8}$&0.6-1.0&2.5 $\pm$ 0.5&7 $>$ 7&BD 60$^\circ$ 39&O9V\\
80&DBSB 48&792$_{ -416}^{+1126}$&45.2$_{ -12.2}^{+11.8}$&1.1&5.0 $\pm$ 0.7&5 $>$ 10&-&O5V\\
81&NGC 2362$^{\dagger}$&809$_{ -409}^{+823}$&29.5$_{ -9.5}^{+30.5}$&3.0&1.39 $\pm$ 0.2&353 $>$ 0.5&$\tau$ CMa&O9Ib\\
82&$[$BDSB2003$]$ 164$^{\ast}$& 842$_{ -429}^{+1065}$&32.2$_{ -8.2}^{+20.8}$&4&3.2&15 $\ge$ 5 $M_\odot$&-& O5-9V\\
83&Pismis 11$^{\dagger}$&896$_{ -448}^{+938}$&40.0$_{ -0.0}^{+40.0}$&3-5&3.6$_{-0.4}^{+0.6}$&43 $>$ 2.5&HD 80077&B2Ia\\
84&$[$FSR2007$]$ 777$^{\ast}$& 949$_{ -758}^{+2166}$&17 $\pm$ 5&3 $\pm$ 2&2.69 $\pm$ 0.3&37 $\ge$ 2.9 $M_\odot$&-&-\\
85&NGC 6530$^{a,\dagger}$&1118$_{ -564}^{+1132}$&55.5$_{ -12.5}^{+13.5}$&2.3&1.35 $\pm$ 0.2&620 $>$ 0.4&9 Sgr&O4V\\
86&$[$FSR2007$]$ 734$^{\ast}$&1175$_{ -833}^{+1202}$&95 $\pm$ 30&2 $\pm$ 1&2.62 $\pm$ 0.3&1266 $\ge$ 0.18 $M_\odot$&-&-\\
87&$[$FSR2007$]$ 761$^{\ast}$& 1184$_{ -893}^{+2426}$&11$_{ -4}^{+3}$&2 $\pm$ 1&2.54 $\pm$ 0.3&73 $\ge$ 1.3 $M_\odot$&-&-\\
88&$[$DBSB2003$]$ 177$^{\ast}$& 1265$_{ -633}^{+1266}$&55.5$_{ -12.5}^{+13.5}$&1&18&23 $\ge$ 5 $M_\odot$&-&O4V\\
89&FSR 1530&1410$_{ -707}^{+1581}$&30.0 $\pm$ 15.0&$<$4.0&2.75 $\pm$ 0.75&35 $>$ 4&[M81]I-296&-\\
90&$[$DB2000$]$ 52$^{\ast}$ &  1416$_{ -724}^{+1591}$&25.1$_{-8.1}^{+9.9}$& 2 $\pm$ 1& 2&25 $\ge$ 5 $M_\odot$& -& O7-8V\\
91&Pismis 5$^{\ast}$&  1428$_{ -714}^{+1446}$& 13$_{ -3}^{+7}$&5 $\pm$ 4&7.2&103 $\ge$ 1.9 $M_\odot$& -& -\\
92&Berkeley 86$^{\dagger}$&1440$_{ -730}^{+1470}$&22.6$_{ -7.6}^{+9.4}$&3-4&1.7&340 $>$ 0.8&HD 193595&O8V(f)\\
93&CC01$^{\ast}$&  1453$_{ -743}^{+1480}$& 15.0 $\pm$ 5.0&2 $\pm$ 1&3.5&520 $\ge$ 0.6 $M_\odot$& L4&B0V\\
94&NGC 637$^{\dagger}$&1682$_{ -854}^{+1726}$&22.6$_{ -7.6}^{+9.4}$&4.0&2.16&583 $M_{\odot}$ $>$ 1.6&-&$\approx$ O8\\
95&$[$DB2000$]$ 26$^{\ast}$& 1705$_{ -852}^{+1721}$& 50.4$_{ -12.4}^{+12.6}$& 2 $\pm$ 1&10 &  31 $\ge$ 5 $M_\odot$& -& O4-5V\\
96&W5Wb$^{\dagger}$&1734$_{ -874}^{+1757}$&25.0$_{ -8.0}^{+12.0}$&2.0&2.0&300 $>$ 1&BD 60$^\circ$ 586&O7.5V\\
97&Stock 16&1857$_{ -955}^{+2045}$&33.4$_{ -8.4}^{+13.3}$&4.0&1.65&16 $>$ 8&HD 115454&O7.5III\\
98&vdB80$^{\ast}$& 2047$_{ -1024}^{+2058}$& 22$_{ -4}^{+3}$&5 $\pm$ 2&2.1&112 $\ge$ 2.3 $M_\odot$& NSV 2998&B8\\
99&ONC$^{\dagger}$&2124$_{ -1078}^{+2175}$&34.7$_{ - 10.7}^{+13.3}$&$<$1.0&0.414 $\pm$ 0.007&3500 (0.1 - 30)&$\Theta$ Orionis C1&O6Vpe\\
100&RCW 38$^{\dagger}$&2251$_{ -1132}^{+2276}$&39.9$_{ -11.9}^{+13.1}$&$<$1.0&1.7&2000 $>$ 0.25&IRS 2&O5.5V\\
101&Bochum 2&2284$_{ -1302}^{+3523}$&27.5$_{ -7.5}^{+9.5}$&2-4&2.7&4 $>$ 16&BD 00$^\circ$ 1617B&O7V\\
102&$[$BDSB2003$]$ 96$^{\ast}$&  2286$_{ -1150}^{+2300}$& 24$_{ -4}^{+2}$&5 $\pm$ 3&1.4&176 $\ge$ 1.8 $M_\odot$&HD 53623&B1II/III\\
103&Berkeley 59$^{\dagger}$&2310$_{ -1168}^{+2417}$&27.5$_{ -7.5}^{+9.5}$&2.0&1.0&41 $>$ 5&BD 66$^\circ$ 1675&O7V\\
104&IC 1590&2376$_{ -1245}^{+2799}$&30.7$_{ -8.7}^{+10.3}$&3.5&2.9&14 $>$ 10&BD 55$^\circ$ 191&O6.5V\\
105&$[$FSR2007$]$ 817$^{\ast}$& 2386$_{ -2133}^{+3506}$&11$_{ -4}^{+3}$&2 $\pm$ 2&2.3 $\pm$ 0.3& 113  $\ge$ 2.5 $M_\odot$&-&-\\
106&W5E$^{\dagger}$&2614$_{ -1323}^{+2667}$&27.5$_{ -7.5}^{+9.5}$&2.0&2.0&400 - 500 $>$ 1&HD 18326&O7V\\
107&W5Wa$^{\dagger}$&2651$_{ -1338}^{+2690}$&39.4$_{ -8.4}^{+10.6}$&2.0&2.0&400 - 500 $>$ 1&HD 17505&O6.5III\\
108&NGC 1931$^{\dagger}$&3128$_{ -1564}^{+3163}$&22.6$_{ -7.6}^{+9.4}$&4.0&3.086&848 $M_{\odot}$ $>$ 2.39&-&$\approx$ O8\\
109&Danks 2$^{\ast}$&  3573$_{ -1832}^{+3990}$& 39.5$_{ -23.3}^{+41.5}$& 3$_{-1}^{+3}$&3.8 $\pm$ 0.6&27 $\ge$ 9 $M_\odot$&D2-1&O8-B3I\\
110&Mercer 23$^{\ast}$& 3687$_{ -1859}^{+3793}$&100$_{ -20}^{+50}$&3 $\pm$ 1&6.5 $\pm$ 0.3&249 $\ge$ 1.9 $M_\odot$&-&WNL7-8/O6If\\
111&LH 118$^{\dagger}$&3746$_{ -1918}^{+4077}$&45.2$_{ -12.2}^{+11.8}$&3.0&48.5&28 $>$ 9&LH 118-241&O5V\\
112&NGC 2103$^{\dagger}$&3853$_{ -1937}^{+3905}$&85.8$_{ -21.8}^{+34.2}$&1.0&48.5&26 $>$ 10&Sk -71$^\circ$ 51&O2V((f$^{\ast}$))\\
113&$[$FSR2007$]$ 944$^{\ast}$& 4163$_{ -3900}^{+10202}$&10 $\pm$ 4&3 $\pm$ 2&2.42 $\pm$ 0.3&298 $\ge$ 1.9 $M_\odot$&-&-\\
114&$[$BDSB2003$]$ 106$^{\ast}$&  4516$_{  -2260}^{+4685}$& 45.1$_{ -17.1}^{+17.9}$& $<$5& 11&82 $\ge$ 5 $M_\odot$&-&O4-6V\\
115&NGC 7380$^{\dagger}$& 4527$_{ -2290}^{+4611}$&37.3$_{ -9.3}^{+10.7}$&2.0&3.7&42 (6 - 12)&HD 215835&O5.5-6V((f))\\
116&GLIMPSE 30$^{\ast}$& 4528$_{ -2297}^{+4619}$& 90 $\pm$ 30&3.5 $\pm$ 0.5&7.2 $\pm$ 0.9&230 $\ge$ 2.35 $M_\odot$&-&WN6-7\\
117&NGC 6231$^{\dagger}$&4595$_{ -2312}^{+4676}$&42.0$_{ -8.0}^{+41.0}$&1.0&1.6&51 $>$ 7&HD 152248&O7Ib\\
118&RCW 106$^{\dagger}$&4681$_{ -2375}^{+4871}$&35.3$_{ -11.3}^{+12.7}$&2.5&1.1&41 $>$ 8&-&O5.5-6.5V\\
119&IC 1484$^{\ast}$& 4705$_{ -2389}^{+4834}$& 34.7$_{ -10.7}^{13.3}$&1.5&2&386 $\ge$ 1.7 $M_\odot$&HD 17505&O6Ve\\
120&NGC 6823&4983$_{ -2584}^{+5685}$&27.5$_{ -7.5}^{+9.5}$&2-4&1.9&42 $>$ 8&BD 22$^\circ$ 3782&O7V(f)\\
121&$[$FSR2007$]$ 846$^{\ast}$&5132$_{ -4791}^{+11621}$&19 $\pm$ 5&3 $\pm$ 2&2.48 $\pm$ 0.3&281 $\ge$ 2.3 $M_\odot$&-&-\\
122&RCW 121$^{\dagger}$&5323$_{ -2671}^{+5390}$&40.0$_{ -12.0}^{+13.0}$&4.2&1.6&96 $>$ 5&-&O5-6V\\
123&$[$FSR2007$]$ 890$^{\ast}$&5345$_{ -5126}^{+13802}$&11$_{-4}^{+3}$&3 $\pm$ 2&2.58 $\pm$ 0.3&288 $\ge$ 2.3 $M_\odot$&-&-\\
 124&$[$FSR2007$]$ 888$^{\ast}$&5935$_{ -5763}^{+16537}$&11$_{-4}^{+3}$&3 $\pm$ 2&2.65 $\pm$ 0.3&249 $\ge$ 2.7 $M_\odot$&-&-\\
125&NGC 2244$^{\dagger}$&5946$_{ -3029}^{+6102}$&55.5$_{ -12.5}^{+13.5}$&1.9&1.5&54 (6 - 12)&HD 46223&O4V(f)\\
126&NGC 2122$^{\dagger}$&6764$_{ -3416}^{+6960}$&46.6$_{ -9.6}^{+30.4}$&3.0&48.5&52 $>$ 9&HD 270145 & O6I(f)\\
127&$[$BDSB2003$]$107$^{\ast}$&  6859$_{-3519}^{+12355}$& 50.4$_{ -12.4}^{+12.6}$&  $<$5&5.8&125 $\ge$ 5 $M_\odot$&-&O4-5V\\
128&Danks 1$^{\ast}$&6952$_{ -3479}^{+6986}$&120$_{ -20}^{+30}$&1.5$_{-0.5}^{+1.5}$&3.8 $\pm$ 0.6&48 $\ge$ 10 $M_\odot$&D1-1&WNLh\\
129&$[$DBS 2003$]$ 179$^{\dagger}$& 7000 $\pm$ 3500&70.0 $\pm$ 10.0&2-5&7.9&10 $>$ 16&Obj 4&Ofpe/WN9\\
130&Westerlund 2$^{\dagger}$&8845$_{ -4456}^{+9009}$&121.0$_{ -43.8}^{+29.0}$&1-3&4.16&29 $>$ 16.5&WR20a A&WN6ha\\
131&RCW 95$^{\dagger}$&9670$_{ -4840}^{+9720}$&67.3$_{ -14.3}^{+15.7}$& 1.5&2.4&136 $>$ 6&-&O3V\\
132&IC 1805$^{\dagger}$&10885$_{ -5528}^{+11137}$&57.0$_{ -11.0}^{+13.0}$ &2.0&2.35&99 (6 - 12)&HD 15558&O4-5III(f)\\
133&NGC 6357&11978$_{ -6430}^{+11979}$&65.8$_{ -7.8}^{+47.2}$&1.0&2.56&38 $>$ 16.5&HDE 319718A&O3If\\
134&NGC 3603$^{\dagger}$&$1.3 \cdot 10^{4}$ $\pm$ 3000&121$_{ -41}^{+29}$& 0.7&6.0&-&NGC 3603-B&WN6ha\\
135&Trumpler 14/16$^{\dagger}$&17890$_{ -8945}^{+18676}$& 99.8$_{ -39.8}^{+50.2}$& 1.7&2.5&64 $>$ 16&$\eta$ Carina &LBV\\
136&NGC 6611$^{\dagger}$&25310$_{ -12659}^{+25503}$&61.7$_{ -10.7}^{+13.3}$&1.3&1.8&460 $>$ 5&HD 168076&O4III\\
137&Cyg OB2$^{\dagger}$&75890$_{ -38453}^{+78716}$&92$_{ -25}^{+58}$&2.0&1.7&8600 $>$ 1.3&Cyg OB2-12&B8Ia\\
138&Arches$^{\dagger}$&77225$_{ -39250}^{+77225}$&111$_{ -41}^{+39}$&2.5&7.62&196 $>$ 20.1&N4&WN7-8h\\
139&R 136$^{\dagger}$&222912$_{ -112104}^{+224426}$&125.4$_{ -45.4}^{+24.6}$& 1-2&48.5&8000 $>$ 3&R136a1&O2If$^\ast$/WN4.5\\
\end{longtable}

\noindent
${\ast}$ These clusters were not included in \citet{WKB09}.\\
${\dagger}$ These clusters constitute the 'low-error' sub-sample (see \S~\ref{sub:stat} for details).\\
${a}$ NGC 6530 was already included in \citet{WKB09} but due to an error the wrong most-massive star was assigned to the cluster. This error does not change any of the \citet{WKB09} results.\\

\clearpage
\begin{longtable}{ccc}
\caption{\label{tab:newclustersref} References for the cluster data in Tab.~\ref{tab:newclusters} in the same order.}\\
\hline
Nr.&Designation&References\\
\hline
\endfirsthead
\caption{continued.}\\
\hline
Nr.&Designation&References\\
\hline
\endhead
\hline
\endfoot
1&IC 348 2&\citet{KM10}\\
2&Cha I 3&\citet{KM10}\\
3&B59&\citet{CLR10}\\
4&Taurus-Auriga 7&\citet{KM10}\\
5&Taurus-Auriga 8&\citet{KM10}\\
6&Taurus-Auriga 6&\citet{KM10}\\
7&Taurus-Auriga 3&\citet{KM10}\\
8&IRAS 05274+3345&\citet{CSS90,CSS93}\\
9&Taurus-Auriga 5&\citet{KM10}\\
10&Mol 139&\citet{FMT09}\\
11&Taurus-Auriga 2&\citet{KM10}\\
12&Taurus-Auriga 4&\citet{KM10}\\
13&Lupus 3&\citet{KM10}\\
14&Cha I 2&\citet{KM10}\\
15&Mol 143&\citet{FMT09}\\
16&Taurus-Auriga 1&\citet{KM10}\\
17&IRAS 06308+0402&\citet{CSS90,CSS93}\\
18&VV Ser&\citet{TPP97,TPN98,TPN99,WL07}\\
19&VY Mon&\citet{TPP97,TPN98,TPN99,WL07}\\
20&Mol 8A&\citet{FMT09}\\
21&IRAS 05377+3548&\citet{CSS90,CSS93}\\
22&Ser SVS2&\citet{KOB04}\\
23&IRAS 05553+1631&\citet{CSS90,CSS93}\\
24&IRAS 05490+2658&\citet{CSS90,CSS93}\\
25&IRAS 03064+5638&\citet{CSS90,CSS93}\\
26&IRAS 06155+2319&\citet{CSS90,CSS93}\\
27&Mol 50&\citet{FMT09}\\
28&Cha I 1& \citet{KM10}\\
29&Mol 11&\citet{FMT09}\\
30&IRAS 06058+2138&\citet{CSS90,CSS93}\\
31&NGC 2023&\citet{Se83,DLG90,LDE91}\\
32&Mol 3&\citet{FMT09}\\
33&Mol 160&\citet{FMT09}\\
34&NGC 7129&\citet{GMM04,WL07}\\
35&IRAS 06068+2030&\citet{CSS90,CSS93}\\
36&IRAS 00494+5617&\citet{CSS90,CSS93}\\
37&V921 Sco&\citet{GMM04,WL07}\\
38&IRAS 05197+3355&\citet{CSS90,CSS93}\\
39&IRAS 05375+3540&\citet{CSS90,CSS93}\\
40&IRAS 02593+6016&\citet{CSS90,CSS93}\\
41&Mol 103&\citet{FMT09}\\
42&NGC 2071&\citet{LDE91}\\
43&Cha I (whole field)&\citet{Lu08,KM10}\\
44&MWC 297&\citet{TPP97,TPN98,TPN99,WL07}\\
45&BD 40$^\circ$ 4124&\citet{TPP97,TPN98,TPN99,WL07}\\
46&$\rho$ Oph&\citet{WLY89,La03,WGA08}\\
47&IRAS 06056+2131&\citet{CSS90,CSS93}\\
48&IRAS 05100+3723&\citet{CSS90,CSS93}\\
49&R CrA&\citet{NF08}\\
50&NGC 1333&\citet{AC03,GFT02,GMM08}\\
51&Mol 28&\citet{FMT09}\\
52&IRAS 02575+6017&\citet{CSS90,CSS93}\\
53&Taurus-Auriga (whole field)&\citet{KM10}\\
54&IC 348 1&\citet{PZ01,LL03,KM10}\\
55&W40&\citet{SBC85,RR08}\\
56&$\sigma$ Ori&\citet{SWW04,BHM08}\\
57&NGC 2068&\citet{Se83,LDE91}\\
58&NGC 2384&\citet{PBM89}\\
59&LkH$\alpha$ 101&\citet{AW08,WWB10}\\
60&Mon R2&\citet{CMD97,PBS02}\\
61&IRAS 06073+1249&\citet{CSS90,CSS93}\\
62&Trumpler 24&\citet{HW84,PBM89}\\
63&IC 5146&\citet{Wa59,FO84,PBM89}\\
&&\citet{WL07,MNL07}\\
64&HD 52266&\citet{DTP04,DTP05}\\
65&HD 57682&\citet{DTP04,DTP05}\\
66&Alicante 5&\citet{MN08}\\
67&Cep OB3b&\citet{NF99,PNJ03,MNL07}\\
68&HD 153426&\citet{DTP04,DTP05}\\
69&Sh2-294&\citet{YDD08}\\
70&NGC 2264&\citet{SBC04,MNL07,Da08}\\
71&RCW 116B&\citet{R07}\\
72&Alicante 1&\citet{NM08}\\
73&RCW 36&\citet{BNN04,BSN06}\\
74&HD 52533&\citet{DTP04,DTP05}\\
75&Sh2-128&\citet{BT03}\\
76&NGC 6383&\citet{RDG03,PNZ07,RD08}\\
77&NGC 2024&\citet{LDE91,HLL00,SWW04}\\
78&HD 195592&\citet{DTP04,DTP05}\\
79&Sh2-173&\citet{CRO09}\\
80&DBSB 48&\citet{OBB08}\\
81&NGC 2362&\citet{MNL07,DH07}\\
82&$[$BDSB2003$]$ 164&\citet{BIM06}\\
83&Pismis 11&\citet{MN08}\\
84&$[$FSR2007$]$ 777&\citet{FSR07}\\
85&NGC 6530&\citet{PDM04,DFM04,DPM06}\\
&&\citet{MNL07,CGZ07}\\
86&$[$FSR2007$]$ 734&\citet{FSR07}\\
87&$[$FSR2007$]$ 761&\citet{FSR07}\\
88&$[$DBSB2003$]$ 177&\citet{BIM05}\\
89&FSR 1530&\citet{FMS08}\\
90&$[$DB2000$]$ 52&\citet{BIM05}\\
91&Pismis 5&\citet{BB09b}\\
92&Berkeley 86&\citet{MJD95,VRC99}\\
93&CC01&\citet{BPI03}\\
94&NGC 637&\citet{HHS08}\\
95&$[$DB2000$]$ 26&\citet{BIM05}\\
96&W5Wb&\citet{KAG08}\\
97&Stock 16&\citet{T85,PBM89}\\
98&vdB80&\citet{BB09b}\\
99&ONC&\citet{HH98,HSC98}\\
&&\citet{MRF07,KWB08}\\
100&RCW 38&\citet{WSB06,WBV08}\\
101&Bochum 2&\citet{Da99,LPS02}\\
102&$[$BDSB2003$]$ 96&\citet{BB09b}\\
103&Berkeley 59&\citet{BW59,Mc68,PSO08}\\
104&IC 1590&\citet{W73,GT97,LPS02}\\
105&$[$FSR2007$]$ 817&\citet{FSR07}\\
106&W5E&\citet{KAG08}\\
107&W5Wa&\citet{KAG08}\\
108&NGC 1931&\citet{HHS08}\\
109&Danks 2&\citet{DCT12}\\
110&Mercer 23&\citet{HKB10}\\
111&LH 118&\citet{MSG89}\\
112&NGC 2103&\citet{MHW05}\\
113&$[$FSR2007$]$ 944&\citet{FSR07}\\
114&$[$BDSB2003$]$ 106&\citet{BIM05}\\
115&NGC 7380&\citet{WSD07}\\
116&GLIMPSE 30&\citet{KBG07}\\
117&NGC 6231&\citet{GM01,SGR06,SRS07,SGN07}\\
118&RCW 106&\citet{RAL03,R07}\\
119&IC 1484&\citet{S55,VHV79,GS92}\\
&&\citet{TWM04,PSK08}\\
120&NGC 6823&\citet{PKK00,LPS02}\\
121&$[$FSR2007$]$ 846&\citet{FSR07}\\
122&RCW 121&\citet{RA06,R07}\\
123&$[$FSR2007$]$ 890&\citet{FSR07}\\
124&$[$FSR2007$]$ 888&\citet{FSR07}\\
125&NGC 2244&\citet{MJD95,PS02,CGZ07}\\
&&\citet{WSD07,BB09}\\
126&NGC 2122&\citet{SM86,GW87}\\
&&\citet{MSG89,NG04}\\
127&$[$BDSB2003$]$107&\citet{BIM05}\\
128&Danks 1&\citet{DCT12}\\
129&$[$DBS 2003$]$ 179&\citet{BIH08}\\
130&Westerlund 2&\citet{BSU04,NRM08,VKB13}\\
131&RCW 95&\citet{RA04b,R07}\\
132&IC 1805&\citet{WSD07}\\
133&NGC 6357&\citet{BTR04,MWM07,WTF07}\\
134&NGC 3603&\citet{HY07,HEM07,SMS08}\\
135&Trumpler 14/16&\citet{PGH93,MJ93,MJD95}\\
&&\citet{NWW04,OC05,AAV07}\\
&&\citet{SCO07,OBB08}\\
136&NGC 6611&\citet{BSS06,WSD07}\\
137&Cyg OB2&\citet{K00,MDW01}\\
&&\citet{WSD07,NMH08}\\
138&Arches&\citet{FNG02,MHP08}\\
139&R 136&\citet{MH98,SMB99,SCC09}\\
\end{longtable}
\end{landscape}


\section{The canonical IMF}
\label{app:IMF}
The following two-component power-law stellar IMF is used throughout the paper:

{\small
\begin{equation}
\xi(m) = k \left\{\begin{array}{ll}
k^{'}\left(\frac{m}{m_{\rm H}} \right)^{-\alpha_{0}}&\hspace{-0.25cm},m_{\rm
  low} \le m < m_{\rm H},\\
\left(\frac{m}{m_{\rm H}} \right)^{-\alpha_{1}}&\hspace{-0.25cm},m_{\rm
  H} \le m < m_{0},\\
\left(\frac{m_{0}}{m_{\rm H}} \right)^{-\alpha_{1}}
  \left(\frac{m}{m_{0}} \right)^{-\alpha_{2}}&\hspace{-0.25cm},m_{0}
  \le m < m_\mathrm{max},\\ 
\end{array} \right. 
\label{eq:4pow}
\end{equation}
\noindent with exponents
\begin{equation}
          \begin{array}{l@{\quad\quad,\quad}l}
\alpha_0 = +0.30&m_\mathrm{low} = 0.01 \le m/{M}_\odot < m_\mathrm{H} = 0.08,\\
\alpha_1 = +1.30&0.08 \le m/{M}_\odot < 0.50,\\
\alpha_2 = +2.35&0.50 \le m/{M}_\odot \le m_\mathrm{max}.\\
          \end{array}
\label{eq:imf}
\end{equation}}
\noindent where $dN = \xi(m)\,dm$ is the number of stars in the mass interval $m$ to $m + dm$. The exponents $\alpha_{\rm i}$ represent the standard or canonical IMF \citep{Kr01,Kr02,KWP13}. For a numerically practical formulation see \citet{PAK06}.

The advantages of such a multi-part power-law description are the easy integrability and, more importantly, that {\it different parts of the IMF can be changed readily without affecting other parts}. Note that this form is a two-part power-law in the stellar regime, and that brown dwarfs contribute about 4 per cent by mass only and that brown dwarfs are a separate population \citep[$k^{'} \approx \frac{1}{3}$,][]{TK07,TK08}.

The observed IMF is today understood to be an invariant Salpeter/Massey power-law slope \citep{Sal55,Mass03} above $0.5\,M_\odot$, being independent of the cluster density and metallicity for metallicities $Z \ge 0.002$ \citep{MH98,SND00,SND02,PaZa01,Mass98,Mass02,Mass03,WGH02,BMK03,PBK04,PAK06}.  Furthermore, un-resolved multiple stars in the young star clusters are not able to mask a significantly different slope for massive stars \citep{MA08,WK07c}. \citet{Kr02} has shown that there are no trends with present-day physical conditions and that the distribution of measured high-mass slopes, $\alpha_3$, is Gaussian about the Salpeter value thus allowing us to assume for now that the stellar IMF is invariant and universal in each cluster. There is evidence of a maximal mass for stars \citep[$m_{\rm max*}\,\approx\,150\,M_{\odot}$,][]{WK04}, a result later confirmed by several independent studies \citep{OC05,Fi05,Ko06}. However, according to \citet{CSH10} $m_\mathrm{max*}$ may also be as  high as 300 $M_\odot$ \citep[but see][]{BK12}. \citet{DKP12} and \citet{MKD10} uncovered a systematic trend towards top-heaviness (decreasing $\alpha_3$) with increasing star-forming cloud density.


\end{appendix}

\bibliography{mybiblio}

\bsp
\label{lastpage}
\end{document}